\documentclass[lettersize,journal]{IEEEtran}
\usepackage{amsmath,amsfonts,amssymb}
\usepackage{graphicx}
\usepackage{subfigure}
\usepackage[colorlinks]{hyperref}
\usepackage{multirow,multicol}
\usepackage{makecell}

\begin{document}
%\title{FAS Empowering 5G NR}
\title{Fluid Antenna System Empowering 5G NR}

\author{Hanjiang Hong,~\IEEEmembership{Member,~IEEE}, 
        Kai-Kit Wong,~\IEEEmembership{Fellow,~IEEE},\\ 
        Haoyang Li,~\IEEEmembership{Member,~IEEE},
        Hao Xu,~\IEEEmembership{Senior Member,~IEEE},
        Han Xiao,~\IEEEmembership{Student Member,~IEEE},\\
        Hyundong Shin,~\IEEEmembership{Fellow,~IEEE}, 
        Kin-Fai Tong,~\IEEEmembership{Fellow,~IEEE}, and
        Yangyang Zhang
\vspace{-6mm}

\thanks{The work of K. K. Wong is supported by the Engineering and Physical Sciences Research Council (EPSRC) under Grant EP/W026813/1.}
\thanks{The work of H. Hong is supported by the Outstanding Doctoral Graduates Development Scholarship of Shanghai Jiao Tong University.}
\thanks{The work of K. F. Tong is supported in part by the Hong Kong Metropolitan University (Staff Research Startup Fund: FRSF/2024/03).}
\thanks{H. Hong and K. K. Wong are with the Department of Electronic and Electrical Engineering, University College London, London, United Kingdom. K. K. Wong is also affiliated with the Department of Electronic Engineering, Kyung Hee University, Yongin-si, Gyeonggi-do 17104, Republic of Korea.}
\thanks{H. Li is with the Cooperative Medianet Innovation Center (CMIC), Shanghai Jiao Tong University, Shanghai 200240, China.}
\thanks{H. Xu is with the National Mobile Communications Research Laboratory, Southeast University, Nanjing 210096, China.}
\thanks{H. Xiao is with the School of Information and Communications Engineering, Xi'an Jiao Tong University, China.}
\thanks{H. Shin is with the Department of Electronics and Information Convergence Engineering, Kyung Hee University, Yongin-si, Gyeonggi-do 17104, Republic of Korea.}
\thanks{K. F. Tong is with the School of Science and Technology, Hong Kong Metropolitan University, Hong Kong SAR, China.}
\thanks{Y. Zhang is with Kuang-Chi Science Limited, Hong Kong SAR, China.}

\thanks{Corresponding author: Kai-Kit Wong.}
}
\maketitle

\begin{abstract}
Fluid antenna system (FAS) is an emerging technology that uses the new form of shape- and position-reconfigurable antennas to empower the physical layer for wireless communications. Prior studies on FAS were however limited to narrowband channels. Motivated by this, this paper addresses the integration of FAS in the fifth generation (5G) orthogonal frequency division multiplexing (OFDM) framework to address the challenges posed by wideband communications. We propose the framework of the wideband FAS OFDM system that includes a novel port selection matrix. Then we derive the achievable rate expression and design the adaptive modulation and coding (AMC) scheme based on the rate. Extensive link-level simulation results demonstrate striking improvements of FAS in the wideband channels, underscoring the potential of FAS in future wireless communications.
\end{abstract}


\section{Introduction}
\IEEEPARstart{W}{ith} the development of fifth generation (5G) wireless communication system in full swing, the development of sixth generation (6G) mobile system is beginning to come into the limelight \cite{Ericsson6G,Andrews20246G}. The vision of 6G is ambitious and it aims to create a seamless reality where the digital and physical worlds merge. Achieving this vision requires new technologies to meet ambitious targets, including, for instance, a peak rate of $1~{\rm Tbps}$, an end-to-end latency of $1~{\rm ms}$, and a connection density of $10^7$ devices/km$^2$, etc \cite{8766143,8869705,Tariq-2020,10054381}.

One of the promising candidates emerging in this effort is fluid antenna system (FAS) \cite{wong2020FAS,wong2022bruce,Wang-2024ai,wu2024fluid}, which leverages the spatial diversity fully in a given spatial region to increase the degree-of-freedom (DoF) in the physical layer for a variety of benefits \cite{New2024aTutorial}. Concisely, FAS is a broad concept that utilizes the new form of shape- and position-reconfigurable antennas for wireless communications. In \cite{Lu-2025}, the definition of fluid antenna was explained using electromagnetic theory. It should be noted that recent terminologies such as movable antennas \cite{zhu2024historical}, flexible-position antenna systems \cite{10480333} and pinching antennas \cite{Yang-2025pa} are well within the definition of FAS.

FAS was first introduced by Wong {\em et al.}~in \cite{9131873,wong2021FAS} when the benefits of a position-flexible antenna at a receiver over wireless channels were studied and highlighted. These seminal works were greatly motivated by reconfigurable antennas such as liquid-metal antennas \cite{huang2021liquid,shen2024design,Shamim-2025}, movable arrays \cite{basbug2017design}, metamaterial-based antennas \cite{johnson2015sidelobe,hoang2021computational,Liu-2025arxiv}, radio-frequency (RF) pixel-based antennas \cite{zhang2024pixel} and etc. Clearly, FAS is not restricted to any specific implementation techniques.\footnote{The word `fluid' in FAS is meant to highlight the flexible nature of the antenna but does not imply the use of fluidic materials for antenna.}

Since the early works in \cite{9131873,wong2021FAS}, there have been numerous attempts in advancing the research of FAS. For instance, performance analysis for FAS has been investigated by using an eigenvalue-based channel model in \cite{Khammassi2023}. The diversity order of a FAS receiver channel was examined in \cite{New2023fluid}. Additionally, \cite{Espinosa-2024} developed a block-correlation model to greatly simplify the performance analysis of FAS. Subsequent analyses further extended the analysis to Nakagami fading channels \cite{Vega2023asimple,Vega2023novel}, and $\alpha$-$\mu$ fading channels \cite{Alvim2023on}. In \cite{new2023information}, the authors considered a multiple-input multiple-output (MIMO) channel with FAS at both ends and derived the diversity and multiplexing tradeoff. Channel estimation is a key aspect for FAS. Recent inquiries into channel estimation for FAS channels have acknowledged the significance of spatial correlation \cite{Skouroumounis2023fluid,xu2024channel,zhang2023successive,Xu-2025ce} and spatial oversampling \cite{10751774}. The interest of FAS has also grown into its application for secrecy communication \cite{Tang-2023,Xu-2024pls,Ghadi-2024dec}, its synergy with reconfigurable intelligent surface (RIS) \cite{10539238,Zhu-2025ris,Yao2025RIS,Salem-2025ris}, and its advantages for integrated sensing and communication (ISAC) \cite{Wang-2024isac,zhou2024fasisac,Zou-2024}.

Furthermore, FAS demonstrates great potential for multiuser communications. The idea is that by exploiting antenna position reconfigurability, FAS can access the received signal in which the interference suffers from a deep fade occurred due to the multipath phenomenon. This has led to the concept of fluid antenna multiple access (FAMA) first proposed in \cite{wong2022FAMA}. Thus far, there have been several variants of FAMA including slow \cite{wong2023sFAMA,Xu2024revisiting,Waqar-2023} and coded \cite{hong2024coded,hong2025Downlink}. It is also possible to combine slow FAMA with analogue signal mixing to improve the interference immunity \cite{Wong2024cuma,Wong-2024cuma-ris}. What makes FAMA an attractive approach is that both interference cancellation at the receiver and precoding at the transmitter are not needed.

Despite the progress so far, the above-mentioned researches were focused on narrowband channels. However, in practical environments, delay spread can be significant, necessitating the use of orthogonal frequency-division multiplexing (OFDM). Technically speaking, there are ongoing concerns if FAS could still perform well under OFDM settings, due to the potential variability in characteristics among the subcarriers. 

Motivated by this, this paper aims to integrate FAS into the OFDM framework. In particular, we present a framework for FAS-OFDM that addresses wideband channels and introduce a novel port selection metric specifically designed for this system. Furthermore, an adaptive modulation and coding (AMC) mechanism tailored to this system is proposed. Importantly, the proposed system is contextualized within the 5G new radio (NR) environment, where we conduct link-level simulations to assess the block error rate (BLER) over the 3rd Generation Partnership Project (3GPP) wideband channels.

Our main contributions are summarized as follows:
\begin{itemize}
\item First, we develop a framework of wideband FAS-OFDM systems, complemented by a novel port selection metric specifically designed for this system. This metric efficiently maps the polymorphic channels in the frequency domain to a distinct channel indicator, thereby enabling FAS to judiciously adjust the antenna port.
\item Based on the proposed port selection metric, we derive the bit-interleaved coded modulation (BICM) capacity for the FAS-OFDM system, which quantifies the achievable rate of the system. Additionally, we investigate the AMC mechanism relevant to the FAS OFDM system, utilizing the derived BICM capacity to calculate the average effective signal-to-noise ratio (SNR) for the system.
\item The proposed FAS-OFDM framework, encompassing the port selection metric and AMC mechanism, is integrated into 5G NR. 5G NR numerologies \cite{38214}, channel coding \cite{38212}, and OFDM modulation \cite{38211} are all adopted in alignment with the latest standards. Link-level simulations are conducted based on the tapped delay line (TDL) channel model from 3GPP \cite{38901} in order to provide an accurate performance evaluation for the FAS-OFDM system. 
\end{itemize}

The remainder of the paper is organized as follows. Section \ref{sec:SystemModel} introduces the system model of the wideband FAS-OFDM system. The port selection, BICM capacity, and AMC mechanism of the FAS-OFDM system are illustrated in Section \ref{sec:PSnAMC}. Section \ref{sec:Sim} specifies the implementation details of 5G NR, and presents the link-level simulation results of the wideband FAS-OFDM system. Conclusions are drawn in Section \ref{sec:conclusion}.

{\em Notations:} Scalars are represented by lowercase letters while vectors and matrices are denoted by lowercase and uppercase boldface letters, respectively. Transpose and hermitian operations are denoted by superscript $T$ and $\dag$, respectively. For a complex scalar $x$, $\lvert x \rvert$ and $x^\dag$ denote its modulus and conjugate, respectively. For a set $\boldsymbol{\chi}$, $|\boldsymbol{\chi}|$ represents its cardinality.

\section{System Model}\label{sec:SystemModel}
\begin{figure*}[]
\centering
\includegraphics[width = 0.75\linewidth]{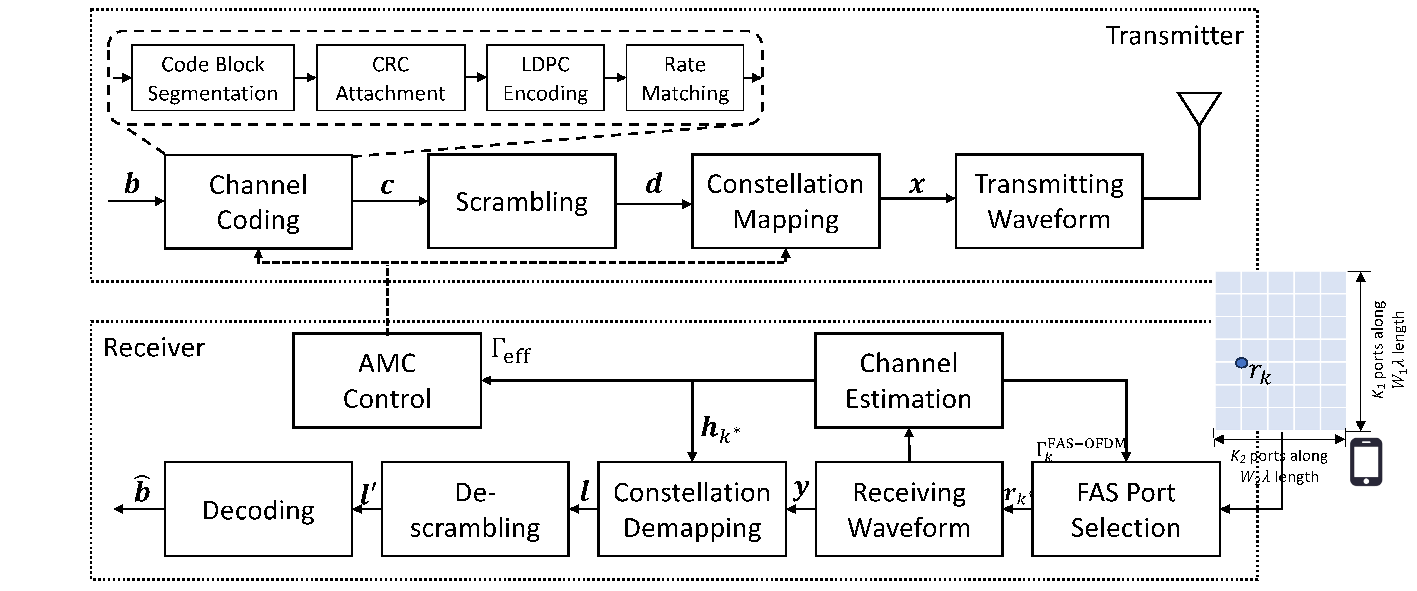}
\caption{System model of the point-to-point FAS communication system.}\label{Fig:SysModel}
\end{figure*}

In this paper, we consider a point-to-point FAS-aided communication system as depicted in Fig.~\ref{Fig:SysModel}. At the transmitter side, the bit stream $\boldsymbol{b} = \{b_0, \dots, b_{N_a-1} \}$ is encoded to a bit sequence $\boldsymbol{c} = \{c_0,\dots, c_{N_b-1} \}$, with a code rate of $\text{CR} = N_a/N_b$. Subsequently, the encoded bit sequence $\boldsymbol{c}$ is scrambled to randomize the date pattern. The resultant scrambled bit sequence $\boldsymbol{d}= \{d_0,\dots, d_{N_b-1}\}$ is then mapped by a constellation $\boldsymbol{\chi}$ into a symbol sequence $\boldsymbol{x} = \{x[0], \dots, x[{N_s-1}]\}$, where $N_s = N_b/Q_m$ represents the length of constellation symbols, and $Q_m = \log_2\lvert\boldsymbol{\chi}\rvert$. Here, normalized constellations are assumed. That is, the average symbol energy is one, or
\begin{equation}
E_x = \frac{1}{\lvert \boldsymbol{\chi} \rvert}\sum_{x\in\boldsymbol{\chi}} \lvert x \rvert ^2 = 1.
\end{equation}
The system can support various code rates and constellations (e.g., quadrature phase shift keying (QPSK), 16 quadrature amplitude modulation (QAM), and 64-QAM, etc.). The symbol sequence $\boldsymbol{x}$ is subsequently used to generate the waveform for transmission. The transmitting waveform generation module may comprise resource mapping and OFDM modulation.

The receiver has a $K=(K_1 \times K_2)$-port two-dimensional FAS (2D-FAS) with a physical size of $W_1\lambda \times W_2 \lambda$, where $\lambda$ denotes the carrier wavelength. Across the 2D space, $K_i$ ports are uniformly positioned along a linear space of length $W_i \lambda$ for $i \in \{1,2\}$. For the sake of simplicity, we define the mapping of the antenna port $(k_1, k_2) \to k: k = k_1\times K_2 +k_2$, where $k_1 \in \{0, \dots, K_1 -1\}$, $k_2 \in \{0, \dots, K_2 -1\}$, and $k \in \{0, \dots, K-1\}$. Utilizing FAS, the received antenna can switch to the port, $k^*$, exhibiting the highest SNR to effectively receive the signal sequence $\boldsymbol{r}_{k^*}$. This sequence is then reversed to form a symbol sequence $\boldsymbol{y} = \{y[0], \dots, y[{N_s-1}]\}$, which is then demapped to the log-likelihood ratio (LLR) sequence $\boldsymbol{l} = \{L_0,\dots, L_{N_b-1}\}$. The soft information $\boldsymbol{l}$ is subsequently passed to the de-scrambler and decoder for the recovery of the data bits $\boldsymbol{\hat{b}} = \{\hat{b}_0, \dots, \hat{b}_{N_a-1} \}$.

\subsection{Narrowband FAS Channel Model}
In the context of narrowband FAS, the $n$-th received symbol at the $k$-th port can be expressed as
\begin{equation}
y_k[n] = h_k x[n] + \eta_k[n],
\end{equation}
where $h_k$ denotes the channel coefficient at the $k$-th port, and $\eta_k[n]$ is the zero-mean complex Gaussian noise with a variance of $\sigma_\eta^2$. The received symbol from all the antenna ports can be consolidated into vector form as
\begin{equation}\label{equ:FASNarrowband}
\boldsymbol{y}[n] = \boldsymbol{h} x[n] + \boldsymbol{\eta}[n],
\end{equation}
where $\boldsymbol{y}[n] = [y_0[n],\dots,y_{K-1}[n]]^T$, $\boldsymbol{h}= [h_0,\dots,h_{K-1}]^T$, and $\boldsymbol{\eta}[n] = [\eta_0[n],\dots,\eta_{K-1}[n]]^T$. We assume that the channels suffer from block fading, which implies that $\boldsymbol{h}$ fluctuates yet remains constant during the transmission of block $\boldsymbol{x}$. In addition, the channels $\{h_k\}_{\forall k}$ exhibit correlation. Utilizing the eigenvalue-based model in \cite{Khammassi2023} and assuming a rich scattering environment, $h_k$ can be expressed as
\begin{equation}
h_k = \sigma_h \sum_{l=1}^{K} \sqrt{\lambda_l}\mu_{k,l}\alpha_l,
\end{equation}
where $\alpha_l \sim \mathcal{CN} (0,1)$, $\lambda_l$ and $\mu_{k,l}$ are derived from the singular value decomposition (SVD) of the channel covariance matrix $\boldsymbol{\Sigma}$, with $\mathbb{E}[\boldsymbol{h} \boldsymbol{h}^\dag] = \sigma_h^2 \boldsymbol{\Sigma}$, and we have 
\begin{equation}\label{Eq:corr}
\left[\boldsymbol{\Sigma}\right]_{k,l}\!=\! J_0\! \left( 2\pi \sqrt{\left(\frac{k_1-l_1}{N_1 - 1} W_1\right)^2 + \left(\frac{k_2-l_2}{N_2 - 1} W_2\right)^2}\right),\end{equation}
where $(l_1, l_2) \to l$, and $J_0(\cdot)$ is the zero-order Bessel function of the first order. SVD is carried out as $\boldsymbol{\Sigma} = \boldsymbol{U}\boldsymbol{\Lambda}\boldsymbol{U}^\dag$, where $\boldsymbol{\Lambda} = {\rm diag} ( \lambda_1, \dots, \lambda_N)$ and $[\boldsymbol{U}]_{k,l} = \mu_{k,l}$.

The received SNR at the $k$-th port is given by 
\begin{equation}
\Gamma_k = \lvert h_k \rvert ^2 \frac{E_x}{\sigma_\eta^2} = \lvert h_k \rvert ^2 \Theta,
\end{equation}
where $\Theta \triangleq \frac{E_x}{\sigma_\eta^2}$. Assuming perfect knowledge of the channel coefficients $\{\lvert h_k \rvert\}_{\forall k}$, the FAS port selection process can be optimized to switch the antenna to the port $k^*$ that provides the maximum received SNR, i.e., 
\begin{equation}\label{Eq:FASPortSelection}
k^* = \arg\max_k | h_k |.
\end{equation}

\subsection{Wideband FAS-OFDM Channel Model}
In this paper, however, the focus is primarily on the wideband FAS-OFDM system, wherein the symbol sequence $\boldsymbol{x}$ is mapped to a symbol grid comprising $N$ OFDM symbols and $F$ subcarriers. The cyclic prefix (CP) eliminates inter-symbol interference (ISI) and accommodates simple frequency-domain processing. At the $n$-th OFDM symbol, the wideband FAS channel can be modelled in the frequency domain as
\begin{equation}\label{equ:FASWideband}
\begin{aligned}
\left[\!\!\! \begin{array}{c} \boldsymbol{y}[0,n] \\ \boldsymbol{y}[1,n] \\ \vdots \\ \boldsymbol{y}[F-1,n] \end{array} \!\!\!\right]  \!=\! & \left[\!\!\! \begin{array}{cccc} \boldsymbol{h}[0,n] \!\!\!& \boldsymbol{0}& \cdots& \boldsymbol{0}\\ \boldsymbol{0}& \!\!\!\boldsymbol{h}[1,n]\!\!\!& \cdots& \boldsymbol{0}\\ \vdots& \vdots& \ddots& \vdots\\ \boldsymbol{0}& \boldsymbol{0}& \cdots& \!\!\!\boldsymbol{h}[F-1,n]\end{array} \!\!\!\right] \\
        & \times\! \left[\!\!\!\begin{array}{c} x[0,n] \\ x[1,n] \\ \vdots \\ x[F-1,n] \end{array}\!\!\!\right] + \left[\!\!\! \begin{array}{c} \boldsymbol{\eta}[0,n] \\ \boldsymbol{\eta}[1,n] \\ \vdots \\ \boldsymbol{\eta}[F-1,n] \end{array}\!\!\!\right],
\end{aligned}
\end{equation}
where $\boldsymbol{y}[f,n] = [y_0[f,n],\dots,y_{K-1}[f,n]]^T$ denotes the received symbol vector, $\boldsymbol{h}[f,n] = [h_0[f,n],\dots,h_{K-1}[f,n]]^T$ denotes the channel vector, $x[f,n]$ denotes the transmitted symbol, and $\boldsymbol{\eta}[f,n] = [\eta_0[f,n],\dots,\eta_{K-1}[f,n]]^T$ denotes the complex noise vector at the $f$-th subcarrier and $n$-th OFDM symbol, respectively, for $0\leq n <N$ and $0\leq f<F$. The channel is subject to frequency-selective fading when $\boldsymbol{h}[f,n]$ varies over $f$, and time-selective fading when $\boldsymbol{h}[f,n]$ varies over $n$. The channels $\{h_k[n,f]\}_{\forall k}$ are also correlated amongst the antenna ports, leading to the covariance matrix as $\mathbb{E}\{\boldsymbol{h}[f,n]\boldsymbol{h}^\dag[f,n]\} = \sigma_h^2\boldsymbol{\Sigma}$, where the elements in $\boldsymbol{\Sigma}$ are specified in \eqref{Eq:corr}.
Notably, when $\boldsymbol{h} [f,n] = \boldsymbol{h}, \forall f \in [0,F), n\in[0,N)$, i.e., the channel is subject to flat fading, the wideband channel model in \eqref{equ:FASWideband} reduces to that expressed in \eqref{equ:FASNarrowband}.

In the wideband FAS-OFDM system, as shown in Fig.~\ref{Fig:SysModel}, the receiving waveform module transfers the received signal from the time domain to the frequency domain. The FAS port selection is proceeded in the time domain, while the channel estimation, demapping, and decoding processes are conducted in the frequency domain. It is impractical for FAS to switch the antenna ports instantaneously in response to frequency-selective fading, as the system cannot select different physical ports for distinct subcarriers. Therefore, the development of a novel port selection metric is imperative to encure the selection of an appropriate port that impacts the performance of multiple subcarriers within the wideband FAS-OFDM system.

\section{Port Selection for Wideband FAS-OFDM}\label{sec:PSnAMC}
In this section, we propose a novel port selection metric for the wideband FAS-OFDM system. The BICM capacity of the system is derived utilizing the proposed metric, and the AMC mechanism is introduced based on the BICM capacity.

\subsection{Port Selection}
Here, we discuss the port selection process, as modeled by \eqref{equ:FASWideband}. Given the assumption of a perfect knowledge of the three-dimensional $K\times N\times F$ channel grid $\boldsymbol{H} = \{\boldsymbol{h}[f,n]\}_{\forall n,f}$, the port selection of FAS-OFDM system requires the formulation of a metric, $\Gamma_k^\text{FAS-OFDM}$, to evaluate the antenna ports and then identify the strongest one (i.e., port $k^*$) by
\begin{equation}\label{Eq:FASOFSMPortSelection}
k^*= \arg\max_k \Gamma_k^\text{FAS-OFDM}.
\end{equation}

This performance metric, $\Gamma_k^\text{FAS-OFDM}$, requires the ability to effectively capture the polymorphic channels $\{h_k[f,n]\}_{\forall f,n}$ and represent them in an alternative single-channel format. Specifically, it can be expressed as
\begin{equation}\label{Eq:snrFASOFDM}
\Gamma_k^\text{FAS-OFDM} = \beta \Psi^{-1}\left(\frac{1}{NF}\sum_{n=0}^{N-1}{\sum_{f=0}^{F-1}{\Psi\left(\frac{\Gamma_k[f,n]}{\beta}\right)}} \right),
\end{equation}
where $\Psi(\cdot)$ represnets a compression function for the SNR mapping, $\Psi^{-1}(\cdot)$ denotes the inverse function of $\Psi(\cdot)$, $\beta$ acts as an adjustment factor, and $\Gamma_k[f,n]$ is the received SNR of $n$-th OFDM symbol and the $f$-th subcarrier, given by 
\begin{equation}
\Gamma_k[f,n] = \lvert h_k[f,n] \rvert ^2 \Theta.
\end{equation}

A crucial component of the performance metric \eqref{Eq:snrFASOFDM} is the design of compression function $\Psi(\cdot)$. Various SNR mapping metrics may be considered, including exponential SNR mapping or mutual information SNR mapping. A straightforward approach is to set $\beta = 1$, and $\Psi(x) = x$, resulting in the average SNR mapping given by
\begin{align}
\Gamma_k^\text{ave} &= \frac{1}{NF}\sum_{n=0}^{N-1}{\sum_{f=0}^{F-1}{\Gamma_k[f,n]}}\notag\\
& = \frac{\Theta}{NF}\sum_{n=0}^{N-1}{\sum_{f=0}^{F-1}{\lvert h_k[h,n] \rvert ^2}}.\label{Eq:aveSNRMap}
\end{align}
By using this average SNR mapping metric, the port selection in \eqref{Eq:FASOFSMPortSelection} can be reformulated as
\begin{equation}\label{Eq:PortSelAVE}
k^*_\text{ave} =\arg\max_k \frac{\Theta}{NF}\sum_{n=0}^{N-1}{\sum_{f=0}^{F-1}{\lvert h_k[h,n] \rvert ^2}}.
\end{equation}
The average SNR in \eqref{Eq:aveSNRMap} corresponds to the carrier-to-noise ratio (CNR). In this case, the port selection can be performed within the frequency domain based on the CNR measurements, leading to a reduction in receiver complexity. 

Note that when the channel is subject to flat fading, i.e., $\boldsymbol{h}[f,n] = \boldsymbol{h}, \forall f,n$, the selection mapping metric in \eqref{Eq:PortSelAVE} degenerates to the FAS port selection metric in \eqref{Eq:FASPortSelection}.

\subsection{Achievable Rate}\label{subsec:AR}
By utilizing the port selection in \eqref{Eq:FASOFSMPortSelection}, the channel grid with the selected $k^*$ port is denoted as $\boldsymbol{H}_\text{FAS} = \left[h_{k^*}[f,n]\right]_{F\times N}$. The received symbol $y_{k^*}[f,n]$ for the $f$-th subcarrier and the $n$-th OFDM symbol can be expressed as
\begin{equation}
    y_{k^*} [f,n] = h_{k^*}[f,n] x[f,n] + \eta_{k^*}[f,n].
\end{equation}

The minimum mean-square-error (MMSE) equalized symbol $\tilde{y}[f,n]$ is calculated as
\begin{align}
\tilde{y}[f,n] & = \beta w[f,n] y_{k^*} [f,n]\notag\\
&= \beta w[f,n] h_{k^*}[f,n] x[f,n] + \beta w[f,n]\eta_{k^*}[f,n],\label{Eq:MMSE}
\end{align}
where $w [f,n]$ represnets the MMSE factor derived from the channel coefficients, expressed as
\begin{equation}
    w [f,n] = \frac{h^\dag_{k^*}[f,n]}{h^\dag_{k^*}[f,n]h_{k^*}[f,n]+\sigma^2_\eta},
\end{equation}
and $\beta$ denotes the normalized factor calculated as
\begin{align}
\beta & = \sqrt{1/\mathbb{E}_{f,n}\left\{\lvert{ w[f,n] h_{k^*}[f,n]}\rvert ^2\right\}}\notag\\
& = \sqrt{\frac{NF}{\sum_{n=0}^{N-1}\sum_{f=0}^{F-1}\lvert w[f,n] h_{k^*}[f,n]\rvert^2}}.
\end{align}
Within the equalized expression \eqref{Eq:MMSE},  the term $\beta w h_{k^*} x$ is the desired signal component, whereas $\beta w \eta_{k^*}$ corresponds to the noise component. The noise power can be found as
\begin{align}
\tilde{\sigma}^2 & = \mathbb{E}_{f,n}\left\{\lvert{\beta w[f,n]\eta_{k^*}[f,n]}\rvert ^2\right\}\notag\\
& = \frac{\beta^2 \sigma_\eta^2}{NF}\sum_{n=0}^{N-1}\sum_{f=0}^{F-1}{\lvert w[f,n]\rvert ^2}.
\end{align}
Thus, the average received SNR $\overline{\Gamma}$ can be formulated as
\begin{align}
\overline{\Gamma} & = \frac{\mathbb{E}_{f,n}\left\{\lvert{\beta w[f,n] h_{k^*}[f,n] x[f,n]}\rvert ^2\right\}}{\mathbb{E}_{f,n}\left\{\lvert{\beta w[f,n]\eta_{k^*}[f,n]}\rvert ^2\right\}}\notag\\
& = \frac{NF\Theta}{\beta^2 \sum_{n=0}^{N-1}\sum_{f=0}^{F-1}\lvert w[f,n]\rvert^2}.
\end{align}

Utilizing this average received SNR $\overline{\Gamma}$, the channel capacity of the FAS-OFDM system can be derived as $C = \log_2(1+\overline{\Gamma})$. More specifically, in the absence of coding and modulation, the system's achievable rate can be evaluated using the average mutual information (AMI, also known as BICM capacity) \cite{BICM}. The BICM capacity $I_{\boldsymbol{\chi}}({\boldsymbol{H}})$ for a given constellation $\boldsymbol{\chi}$ of this channel can be articulated as \eqref{Eq:AMIH} (see top of next page),
\begin{figure*}[]
\begin{equation}\label{Eq:AMIH}
I({\boldsymbol{H}}) = \log_2 \lvert \boldsymbol{\chi} \rvert + \frac{1}{NF}\sum_{n=0}^{N-1} \sum_{f=0}^{F-1} \sum_{i = 1}^{\log_2 \lvert \boldsymbol{\chi} \rvert} \sum_{b=0}^{1}       {{\frac{\sum_{x\in {\boldsymbol{\chi}_i^{(b)}}}\! p(\tilde{y}[f,n]|x,h_{k^*}[f,n])}{\sum_{x\in {\boldsymbol{\chi}}} p (\tilde{y}[f,n]|x,h_{k^*}[f,n])}} \log_2{\frac{\sum_{x\in {\boldsymbol{\chi}_i^{(b)}}}\! p(\tilde{y}[f,n]|x,h_{k^*}[f,n])}{\sum_{x\in {\boldsymbol{\chi}}} p (\tilde{y}[f,n]|x,h_{k^*}[f,n])}}}
\end{equation}
\hrulefill
\end{figure*}
where the probability density function (PDF) $p (\tilde{y}|x,{h_{k^*}})$ is given by
\begin{equation}\label{Eq:pdf}
p(\tilde{y}|x,{h_{k^*}}) = \frac{1}{\pi\tilde{\sigma}^2} \exp\left({-\frac{\lvert \tilde{y}- x \rvert ^2}{\tilde{\sigma}^2}}\right).
\end{equation}
The BICM capacity of the FAS-OFDM system can be computed as the expectation over the channel grid $\boldsymbol{H}$, i.e.,
\begin{equation}
I_{\text{FAS-OFDM},\boldsymbol{\chi}} = \mathbb{E}_{\boldsymbol{H}} \left[I_{\boldsymbol{\chi}}({\boldsymbol{H}})\right].
\end{equation}

When the channel experiencing flat fading, i.e., $h_{k^*}[f,n] = h_{k^*}, \forall f,n$, the BICM capacity can also be derived in integral form as \eqref{Eq:AMI_hint} (see top of next page),
\begin{figure*}[tb]
\begin{equation}\label{Eq:AMI_hint}
I_\text{FAS} = \int_{r=0}^{\infty}I_{\boldsymbol{\chi}}(r) p(r)dr = \log_2 \lvert \boldsymbol{\chi} \rvert \!+\! \frac{1}{\lvert \boldsymbol{\chi} \rvert} \int_{r=0}^{\infty} \iint_D \sum_{i = 1}^{\log_2 \lvert \boldsymbol{\chi} \rvert}\sum_{b=0}^{1} {\sum_{x\in {\boldsymbol{\chi}_i^{(b)}}}\! p(\tilde{y}|x,r) \log_2 {\frac{\sum_{x\in {\boldsymbol{\chi}_i^{(b)}}}\! p(\tilde{y}|x,r)}{\sum_{x\in {\boldsymbol{\chi}}}\! p(\tilde{y}|x,r)}}} p(r) d\tilde{y}dr
\end{equation}
\hrulefill
\end{figure*}
where the PDF of the FAS channel, $p(r) \triangleq p(|h_{k^*}|=r)$, has been given in \cite{wong2021FAS,New2023fluid}.

\subsection{AMC Mechanism}\label{subsec:AMCMech}
The objective of AMC is to allow the FAS-OFDM system to dynamically adjust its constellation and channel code rate based on real-time channel conditions, hence enhancing both reliability and efficiency in communication. 

In an ideal system in which AMC can be seamlessly controlled, modulation and code rates are derived by maximizing the BICM capacity, as articulated in \eqref{Eq:AMIH}. The BICM capacity corresponding to multiple modulation constellations $\{\boldsymbol{\chi}_i\}$ is calculated to identify the optimal constellation $\boldsymbol{\chi}_\text{opt}$, which offers the highest capacity. Hence, the corresponding code rate is then determined as $\text{CR} = I_{\boldsymbol{\chi}_\text{opt}}(\boldsymbol{H})/\log_2|\boldsymbol{\chi}_\text{opt}|$.

Nevertheless, achieving a truly seamless AMC in practice is often challenging. In the pragmatic system, the AMC control process generates feedback in the form of the Channel Quality Index (CQI), which is transmitted back to the transmitter side. The precision of AMC relies heavily on the accuracy of the CQI's bit representation. The implementation process of CSI feedback follows several steps:
\begin{itemize}
\item \textbf{Creation of a look-up table (LUT)}: This table connects SNR to CQI based on computer simulations conducted in the additive white Gaussian noise (AWGN) channel;
\item \textbf{Effective SNR calculation}: The effective SNR, $\Gamma_\text{eff}$, for the FAS-OFDM system is calculated, and this effective SNR is mapped to the approaching CQI value using the pre-defined LUT;
\item \textbf{CQI transmission}: The computed CQI is then sent to the transmitter side, which selects the proper modulation constellation and code rate based on the received CQI.
\end{itemize}

In FAS-OFDM, the effective SNR, $\Gamma_\text{eff}$, can be derived from the inverse function of the BICM capacity $I(\cdot)$ as 
\begin{equation}\label{Eq:effSNR}
\Gamma_\text{eff} = \alpha I^{-1}\left[I(\boldsymbol{H})\right],
\end{equation}
where $I(\boldsymbol{H})$ is the average BICM capacity of the $F\times N$ grid calculated as per \eqref{Eq:AMIH}, and $\alpha$ is the adjustment factor, which can be tailored to correspond to the modulation constellation and code rate of the chosen CQI. The MMSE method is used to calibrate the adjustment factor $\alpha$, given by
\begin{equation}
\alpha_\text{opt} =\arg\min_\alpha \left\{ \sum_{i=1}^{N_c}\lvert \Delta e_i (\alpha)\rvert ^2\right\},
\end{equation}
where
\begin{equation}
\Delta e_i (\alpha) = \log_{10} \Gamma_\text{eff}^{(i)}(\alpha) - \log_{10} \Gamma_\text{LUT}^{(i)},
\end{equation}
where $N_c$ denotes the number of channel realizations, $\Gamma_\text{eff}^{(i)}(\alpha)$ and $\Gamma_\text{LUT}^{(i)}$ are the effective SNR calculated by \eqref{Eq:effSNR} and the SNR from the LUT, respectively.

\section{Application to 5G NR System\\and Simulation Results}\label{sec:Sim}
In this section, we employ the port selection metric and the AMC mechanism discussed in Section \ref{sec:PSnAMC} to 5G NR systems. The transmission of physical data share channel (PDSCH) is considered. In the procedures associated with PDSCH \cite{38214}, the channel coding includes the code block segmentation, the cyclic redundancy check (CRC) attachment, the low-density parity-check (LDPC) encoding, and the rate matching \cite{38212}. The length of uncoded bits in a subframe, $N_a$, corresponds to the transmit block size (TBS), while the length of transmission data bits is determined by $N_b = N_\text{RE} \times Q_m$, with $N_\text{RE} = N_s = N'_\text{RE}\times N_\text{PRB}$ being the number of resource elements allocated for data transmission in a subframe, and $N'_\text{RE}$ is the number of resource elements per physical resource block (PRB). The overall code rate can be expressed as 
\begin{equation}
\text{CR} = N_a/N_b = \text{TBS}/(N_\text{RE}\times Q_m).
\end{equation}
The BICM efficiency is thereby denoted as
\begin{equation}\label{Eq:BICME}
E_\text{BICM} = \text{CR} \times Q_m = \text{TBS}/N_\text{RE}~\text{[bit/channel use]}.
\end{equation}

Following constellation mapping, the resultant symbol sequence is mapped to the PRBs for OFDM transmission \cite{38211}. In this paper, we focus on the FAS performance, operating under the assumptions of static resource allocation and constant power allocation. Thus, the transmission symbol grid $\boldsymbol{X}$ comprises $N = N_\text{symb}^\text{subframe}$ OFDM symbols and $F = N_\text{PRB} \times N_\text{sc}^\text{RB}$ subcarriers, where $N_\text{symb}^\text{subframe}$ is the number of OFDM symbols per subframe, $N_\text{PRB}$ refers to the number of PRBs, and $N_\text{sc}^\text{RB}$ denotes the number of subcarriers per resource block (RB). The symbol matrix $\boldsymbol{X}$ is OFDM modulated to produce a time-domain symbol sequence for transmission. 

The overall system throughput can be calculated as
\begin{equation}\label{Eq:throughput}
\gamma = \text{TBS}/T_\text{subframe} ~\text{[bps]},
\end{equation}
where $T_\text{subframe} = 1~{\rm ms}$ is the duration of a subframe. The overall spectrum efficiency is then calculated as 
\begin{equation}\label{Eq:SE}
E = \gamma/\text{BW} ~\text{[bps/Hz]},
\end{equation}
where $\text{BW}$ denotes the bandwidth. The relationship between the spectrum efficiency in \eqref{Eq:SE} and the BICM efficiency in \eqref{Eq:BICME} can be expressed as
\begin{equation}\label{Eq:rela_E_EBICM}
E = E_\text{BICM} (1-\epsilon_\text{RS})(1-\epsilon_\text{CP})(1-\epsilon_\text{GB}),
\end{equation}
in which $\epsilon_\text{RS}$, $\epsilon_\text{CP}$, $\epsilon_\text{GB}$ account for the overhead of reference signals, CP, guard bands (GB), respectively, and are given by
\begin{equation}\label{Eq:overheads}
\left\{\begin{aligned}
\epsilon_\text{RS} &= 1- \frac{N_\text{RE}}{N_\text{PRB}\times N_\text{sc}^\text{RB}\times N_\text{symb}^\text{subframe}}, \\
\epsilon_\text{CP} &= \frac{T_\text{CP}}{T_\text{subframe}}, \\
\epsilon_\text{GB} &=  1-\frac{\text{ABW}}{\text{BW}}.
\end{aligned}\right.
\end{equation}
In \eqref{Eq:overheads}, $T_\text{CP}$ is the duration of CP, and $\text{ABW} = N_\text{PRB}\times N_\text{sc}^\text{RB}\times \Delta f$ denotes the available bandwidth. % The CP overhead $\epsilon_\text{CP} = 1/15$, considering normal CP in this paper. 

The signals are transmitted over the 3GPP TDL channel \cite{38901} based on the correlation matrix $\boldsymbol{\Sigma}$ in \eqref{Eq:corr}. The power and delay profile for the TDL channel are detailed in \cite[Clause 7.7.2]{38901}. For AMC, CQI is represented by $4$ bits, ranging from $0$ to $15$. Every CQI, aside from CQI 0, corresponds to a specific combination of code rate and modulation constellation. We use \cite[Table 5.2.2.1-2]{38214}, and the combinations of code rate and modulation are outlined in Table \ref{Tab:CQI}. The user equipment (UE) assesses and reports back the CQI to next generation NodeB (gNB), which subsequently selects an modulation and coding scheme (MCS) from the $32$ MCSs defined in \cite[Table 5.1.3.1-1]{38214} for the UE that aligns with the constellation and closely matches the code rate indicated by the CQI.

In the simulations, the system parameters are summarized in Table \ref{Tab:SimPara}. We mainly focus on the system with bandwidth of $5$ MHz, though comparisons with systems utilizing different bandwidths are addressed in Section \ref{subsec:BLERPerf}. In Sections \ref{subsec:BLERPerf} and \ref{subsec:Mob}, we employ static MCS 7 with target code rate of $526/1024$ and QPSK modulation from \cite[Table 5.1.3.1-1]{38214}, while Sections \ref{subsec:amc} and \ref{subsec:SE} evaluate the AMC using the CQI table in \cite[Table 5.2.2.1-2]{38214}. Sections \ref{subsec:BLERPerf}, \ref{subsec:amc}, and \ref{subsec:SE} specifically focus on the scenarios with a short delay spread of $30$ ns and a maximum Doppler frequency of $30$ Hz, while Section \ref{subsec:Mob} assesses the system under varying delay spreads and maximum Doppler frequency profiles.

\begin{table}[t]
\begin{center}
    \vspace{-2mm}
    \caption{Simulation Parameters}
    \label{Tab:SimPara}
    \begin{tabular}{l|l|l|l|l}
        \hline
        \textbf{Parameter}                  & \multicolumn{4}{l}{\textbf{Value}} \\ \hline\hline
        \makecell[l]{Normalized size of FAS \\
            $W_1$ or $W_2$}     & \multicolumn{4}{l}{$0.2$, $0.5$, $1$, $2$, $5$} \\ \hline
        Number of ports $K_1$ or $K_2$    & \multicolumn{4}{l}{From $1$ to $20$}      \\ \hline
        Subcarrier spacing $\Delta f$       & \multicolumn{4}{l}{$15$ kHz}          \\ \hline
        \makecell[l]{Number of subcarriers \\
            per RB $N_\text{sc}^\text{RB}$}    & \multicolumn{4}{l}{$12$}  \\ \hline
        \makecell[l]{Number of symbols \\
            per subframe $N_\text{symb}^\text{subframe}$}  & \multicolumn{4}{l}{$14$}              \\ \hline
            \makecell[l]{Number of resource elements \\
            in a PRB $N'_\text{RE}$} & \multicolumn{4}{l}{$156$} \\ \hline
        Cyclic prefix (CP)                         & \multicolumn{4}{l}{Normal}  \\ \hline
        Bandwidth (BW) [MHz]                     & $1.4$    & $5$   & $10$  & $20$         \\ \hline
        Number of PRBs $N_\text{PRB}$       & $6$   & $25$  & $50$    & $100$              \\ \hline 
        FFT Size $N_\text{fft}$             & $128$ & $512$ & $1024$ 
        &  $2048$            \\ \hline
        Channel Model       & \multicolumn{4}{l}{TDL-C channel} \\ \hline
        Channel Estimation  & \multicolumn{4}{l}{Perfect} \\ \hline
        Delay spread        & \multicolumn{4}{l}{$10$, $30$, $100$, $300$, $1000$ ns} \\ \hline
        Maximum Doppler frequency   & \multicolumn{4}{l}{$0$, $30$, $100$, $300$, $1000$ Hz} \\ \hline
        Demapping Scheme    & \multicolumn{4}{l}{Max-Log-MAP} \\ \hline
        Decoding Scheme     & \multicolumn{4}{l}{Min-Sum}      \\ \hline
    \end{tabular}
\end{center}
\end{table}

\subsection{BLER Performance}\label{subsec:BLERPerf}
\begin{figure}[]
\begin{center}
\subfigure[$W=0.2\lambda\times 0.2\lambda$]{\includegraphics[width=\linewidth]{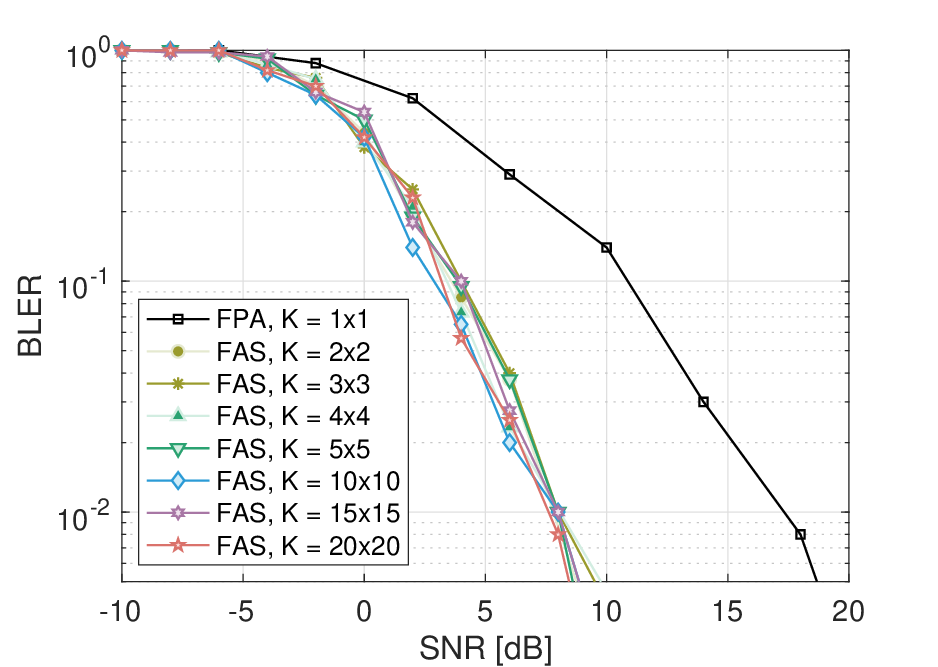}\label{SubFig:BLERvsSNR_W0.2x0.2}}
\subfigure[$W=1\lambda\times 1\lambda$]{\includegraphics[width=\linewidth]{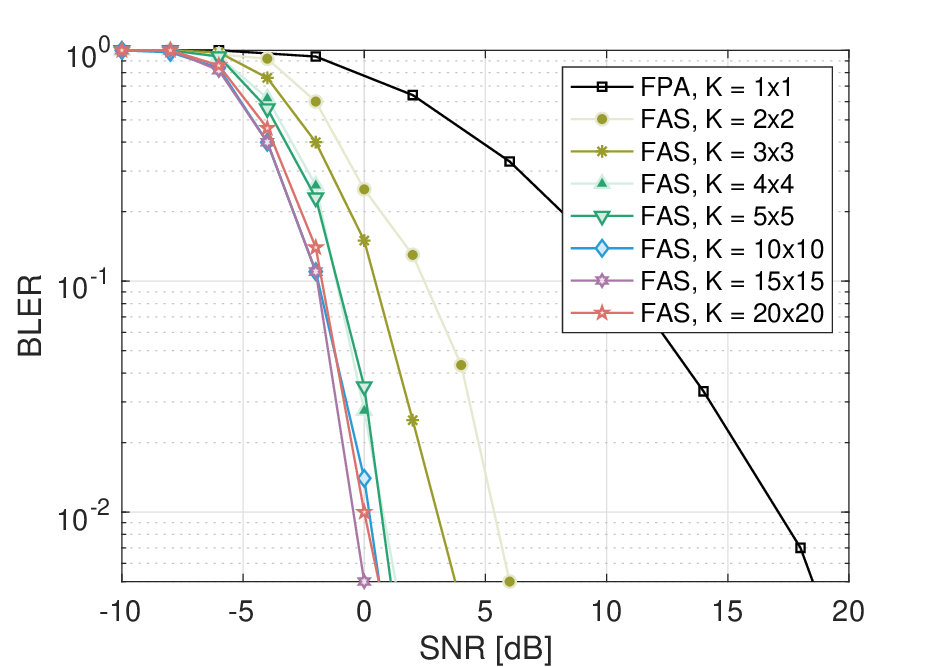}\label{SubFig:BLERvsSNR_W1x1}}
\subfigure[$W=5\lambda\times 5\lambda$]{\includegraphics[width=\linewidth]{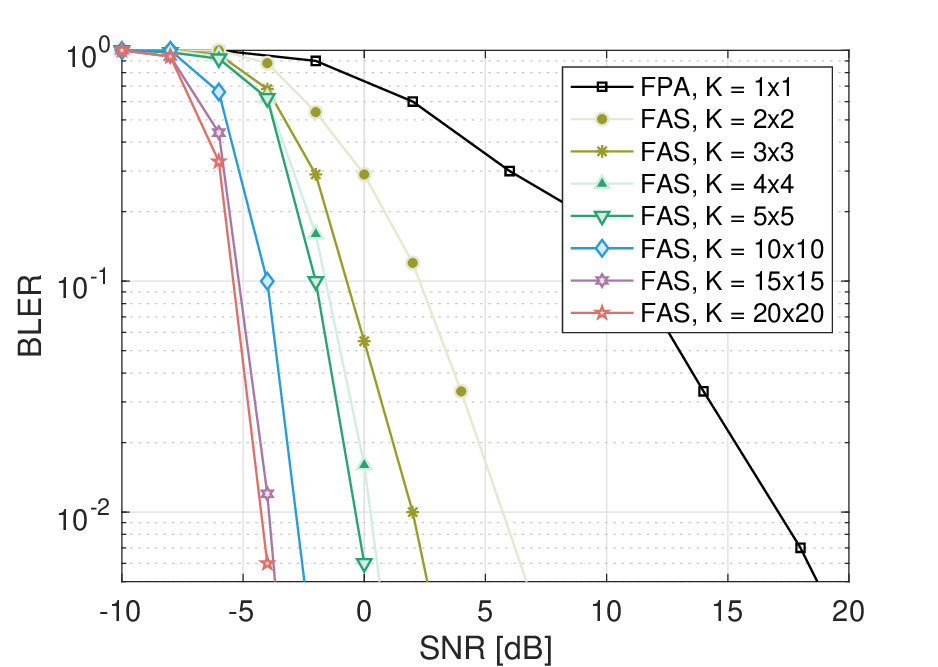}\label{SubFig:BLERvsSNR_W5x5}}
\caption{BLER performance comparison for different number of FAS antenna ports, $K$. Results are presented for MCS7, with FAS physical size of (a) $W = 0.2\lambda \times 0.2\lambda$, (b) $W = 1\lambda \times 1\lambda$, and (c) $W = 5\lambda \times 5\lambda$.}\label{Fig:BLERDiffN}
\end{center}
\end{figure}

Fig.~\ref{Fig:BLERDiffN} demonstrates the BLER performance of the wideband FAS-OFDM system, for a comparative analyses against the number of FAS antenna ports. The BLER of the fixed position antenna (FPA) system with a configuration of $K = 1\times 1$ is also included for comparison. As can be seen, FAS outperforms FPA greatly, even when the FAS has a relatively small size of $W = 0.2\lambda\times 0.2 \lambda$, or very few ports, specifically $K=2\times 2$. When the physical size of FAS, $W$, is small, the performance enhancement through an increase in the number of antenna ports is considerably constrained due to significant correlation among the antenna ports. For instance, when $W = 0.2\lambda \times 0.2\lambda$, as illustrated in Fig.~\ref{Fig:BLERDiffN}\ref{sub@SubFig:BLERvsSNR_W0.2x0.2}, a performance gain of $10$ dB is achieved if $\text{BLER}=10^{-2}$ is to be satisfied, corresponding to an increase in the number of ports from $K = 2\times 2$ to $20\times 20$. However, with an appropriately scaled physical size, BLER performance exhibits improvement as the number of antenna ports escalates until a saturation point is reached, beyond which no further enhancement in performance is observable with additional antenna ports. For example, when the physical size is established as $W = 5\lambda \times 5\lambda$ in Fig.~\ref{Fig:BLERDiffN}\ref{sub@SubFig:BLERvsSNR_W5x5}, a performance gain of $11$ dB is observed when $K=2\times 2$, increasing to $21$ dB for $K=20\times 20$. The number of ports to reach saturation is higher for larger size $W$. In Fig.~\ref{Fig:BLERDiffN}\ref{sub@SubFig:BLERvsSNR_W0.2x0.2}, the point of saturation is identified at $K = 2\times 2$ for $W =0.2\lambda \times 0.2\lambda$, while it extends to approximately $K = 10\times 10$ for $W =1\lambda \times 1\lambda$ in Fig.~\ref{Fig:BLERDiffN}\ref{sub@SubFig:BLERvsSNR_W1x1}, and approaches $K = 20\times 20$ for $W =5\lambda \times 5\lambda$ in Fig.~\ref{Fig:BLERDiffN}\ref{sub@SubFig:BLERvsSNR_W5x5}. Moreover, comparing the FPA results with different physical size, the performance of FPA remains invariant with respect to alterations in the physical size of the antenna.

\begin{figure}[]
\begin{center}
\subfigure[$K=2\times 2$]{\includegraphics[width=\linewidth]{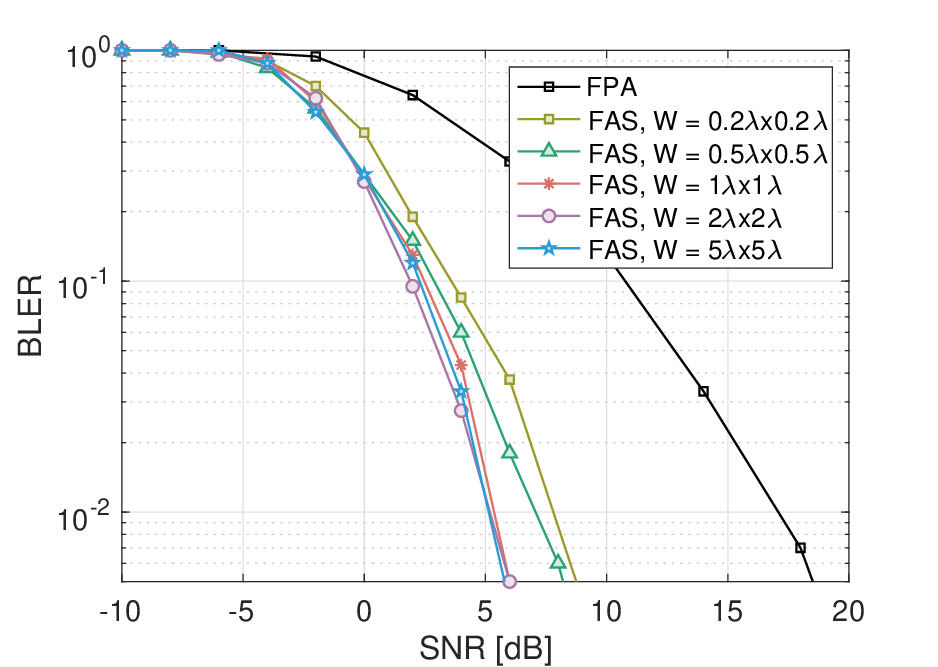}\label{SubFig:BLERvsSNR_N2x2}}
\subfigure[$K=5\times 5$]{\includegraphics[width=\linewidth]{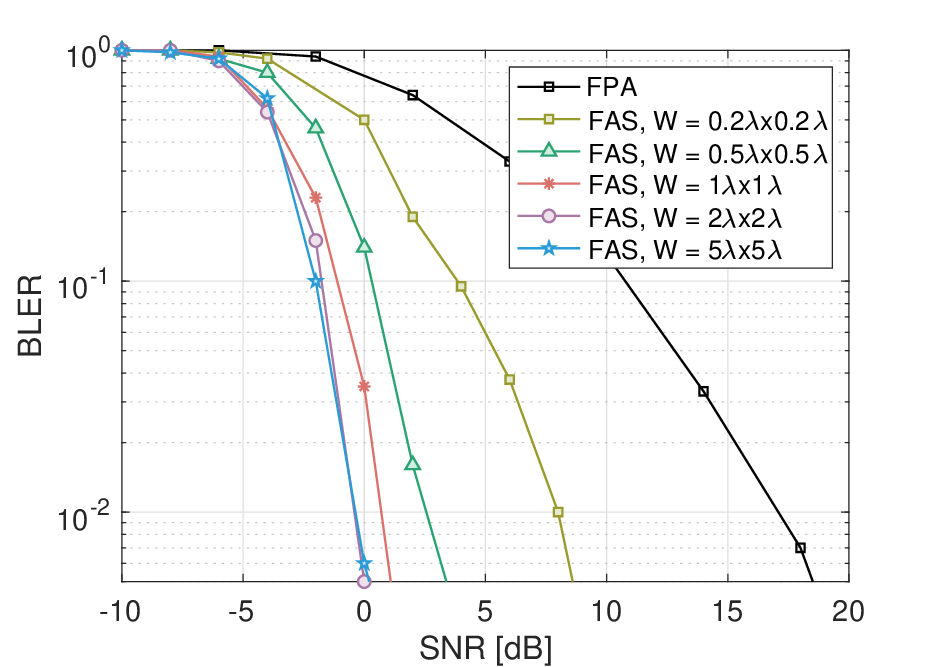}\label{SubFig:BLERvsSNR_N5x5}}
\subfigure[$K=20\times 20$]{\includegraphics[width=\linewidth]{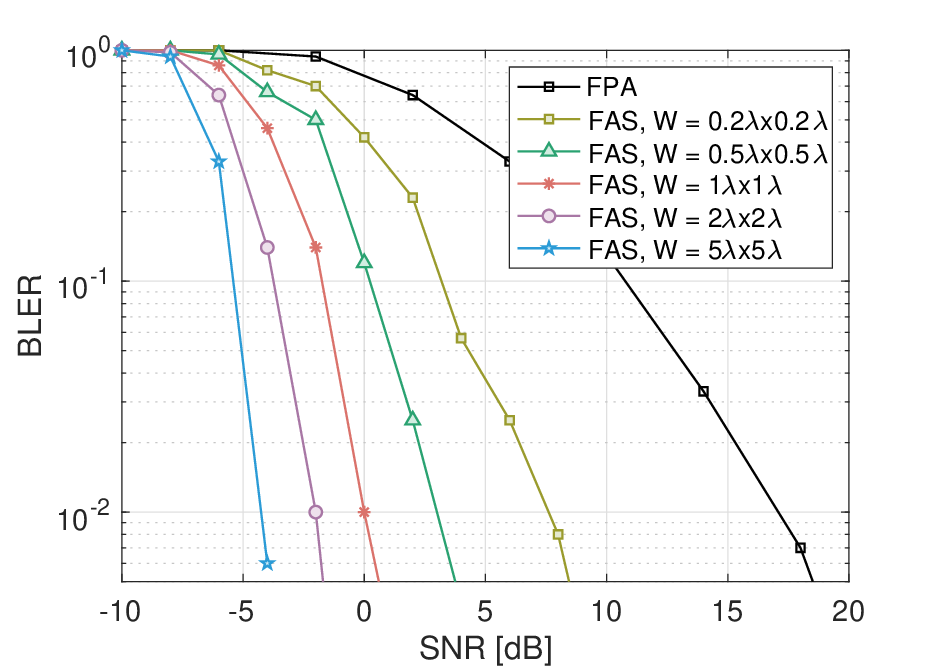}\label{SubFig:BLERvsSNR_N20x20}}
\caption{BLER performance comparison for different size of FAS, $W$. Results are presented for MCS7, with (a) $K = 2\times 2$, (b) $K=5\times 5$, and (c) $K=20\times 20$ antenna ports.}\label{Fig:BLERDiffW}
\end{center}
\end{figure}

In Fig.~\ref{Fig:BLERDiffW}, the BLER results are provided with different FAS physical size $W$. As expected, the performance gain of FAS is markedly higher with larger physical size, particularly when the number of antenna ports $K$ is deemed sufficient. With a limited number of FAS ports, such as $K = 2\times 2$ in Fig. \ref{Fig:BLERDiffW}\ref{sub@SubFig:BLERvsSNR_N2x2}, the BLER performance exhibits an upward trend correlating with the physical size until reaching the saturation point rapidly. At the saturation point, the correlation among the antenna ports becomes negligible, and the performance is predominantly constrained by the limited number of the antenna ports. Conversely, with an adequate number of antenna ports, such as $K = 20\times 20$ in Fig. \ref{Fig:BLERDiffW}\ref{sub@SubFig:BLERvsSNR_N20x20}, FAS demonstrates superior performance gains as the physical size increases. 

\begin{figure}[]
\centering
\subfigure[$W = 1\lambda \times 1\lambda$]{\includegraphics[width=\linewidth]{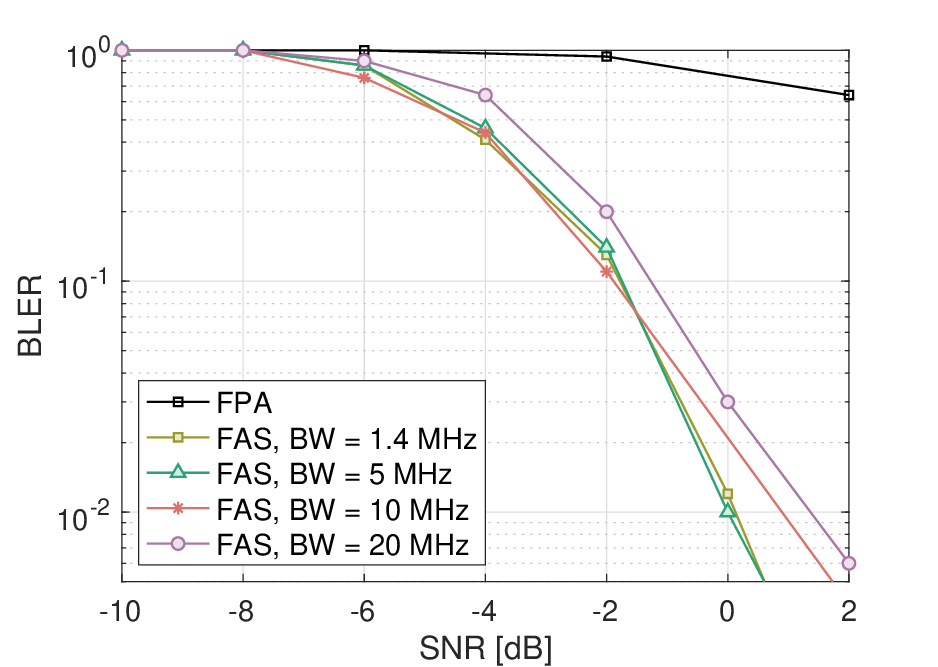}\label{SubFig:BLERvsSNR_DiffBW_W1x1}}
\subfigure[$W = 5\lambda \times 5\lambda$]{\includegraphics[width=\linewidth]{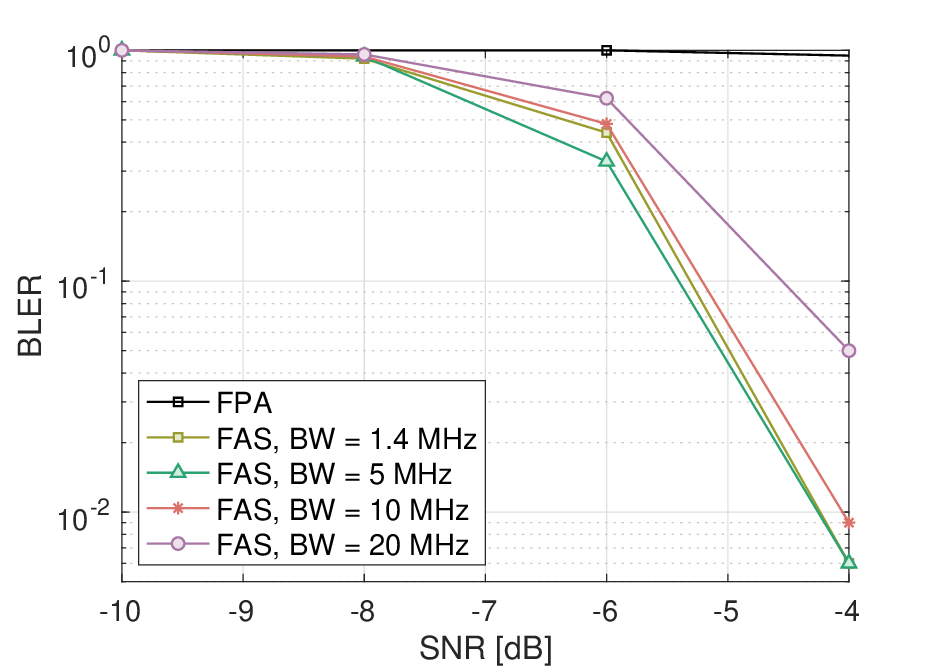}\label{SubFig:BLERvsSNR_DiffBW_W5x5}}
\caption{BLER performance comparison for different bandwidth. Results are presented for MCS7, with $K = 20\times 20$ antenna ports over the FAS with physical size of (a) $W = 1\lambda \times 1\lambda$ and (b) $W = 5\lambda \times 5\lambda$.}\label{Fig:BLERDiffBW}
\end{figure}

The results in Fig.~\ref{Fig:BLERDiffBW} study the BLER in relation to bandwidth. The results indicate a modest performance degradation with increasing bandwidth. We see that FAS-OFDM performs better at narrower bandwidths. This phenomenon is attributed to the reality that systems operating with broader bandwidths incorporate more subcarriers, rendering them more susceptible to frequency-selective fading. In a high-dynamic channel, the subframe average SNR in \eqref{Eq:aveSNRMap} tends to converge toward the average SNR $\Theta$, which may hinder resolution among the fluid antenna ports. Nevertheless, the simulation results indicate that the observed performance degradation remains minor. FAS-OFDM continues to provide considerable performance gains even at a bandwidth of $20$ MHz. Notably, the performance at the bandwidth of $1.4$ MHz is slightly worse than that at $5$ MHz bandwidth. This discrepancy ascribed to the actual spectral efficiency of MCS 7 in the $1.4$ MHz bandwidth system being slightly higher than that in the $5$ MHz bandwidth system, in accordance with the TBS determination of 5G NR.

\subsection{Mobility Evaluation}\label{subsec:Mob}
\begin{figure}[]
\centering
\subfigure[$W = 1\lambda \times 1\lambda$]{\includegraphics[width=\linewidth]{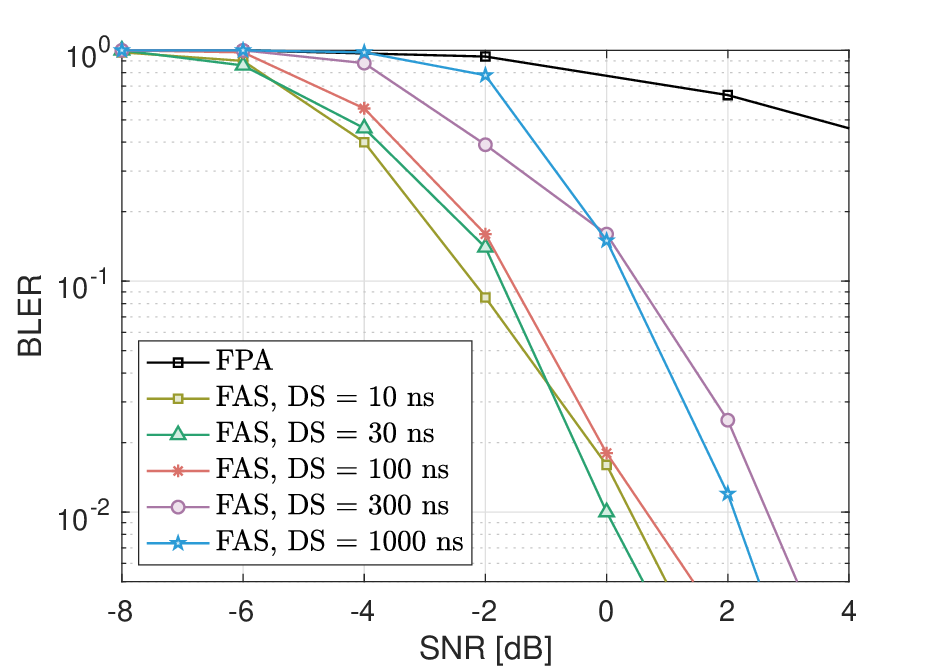}\label{SubFig:BLERvsSNR_DiffDS_W1x1}}
\subfigure[$W = 5\lambda \times 5\lambda$]{\includegraphics[width=\linewidth]{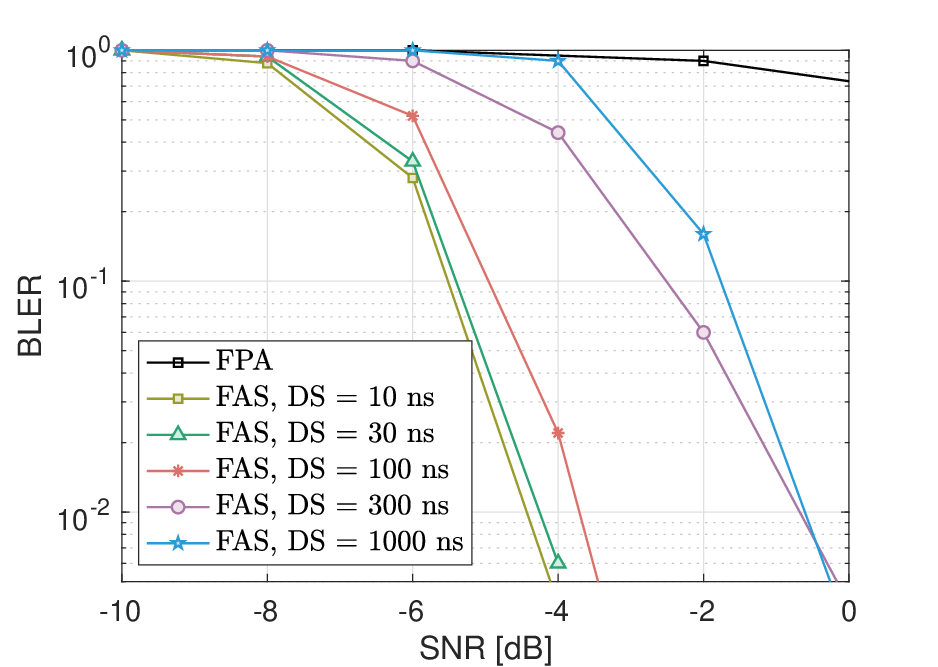}\label{SubFig:BLERvsSNR_DiffDS_W5x5}}
\caption{BLER performance comparison for different delay spread. Results are presented for MCS7, with $K = 20\times 20$ antenna ports over the FAS with physical size of (a) $W = 1\lambda \times 1\lambda$ and (b) $W = 5\lambda \times 5\lambda$.}\label{Fig:BLERDiffDS}
\end{figure}

To evaluate the performance of FAS-OFDM under mobility scenarios, simulations are conducted under time-varying delay spread and maximum Doppler shift profiles. Fig.~\ref{Fig:BLERDiffDS} presents the results associated with different delay spread profiles. Delay spread is closely related to the propagating environment. The adopted delay spread profiles, ranging from $10$ ns to $1000$ ns, are categorized according to \cite[Table 7.7.3-1]{38901} as very short, short, nominal, long, and very long delay spreads, respectively. As can be seen, the BLER performance degrades as the delay spread increases, particularly in the system exhibiting long or very long delay spreads. Performance degradation occurs as the coherence bandwidth is smaller than the system bandwidth. Specifically, for a long delay spread ($\text{DS} = 300$ ns), the coherence bandwidth is $\text{BW}_\text{c} = 3.3$ MHz, while the coherence bandwidth for a very long delay spread ($\text{DS} = 1000$ ns) is $\text{BW}_\text{c} = 1$ MHz, both of which fall below the system bandwidth of $\text{BW} = 5$ MHz. Consequently, frequency-selective fading becomes more severe, leading to increased ISI that distorts the signal. But the performance degradation remains relatively small, amounting to less than $2$ dB or $4$ dB for configurations of $W = 1\lambda \times 1\lambda$ or $5\lambda \times 5\lambda$, respectively. It can still achieve considerable performance gain compared to FPA.

\begin{figure}[]
\centering
\subfigure[$W = 1\lambda \times 1\lambda$]{\includegraphics[width=\linewidth]{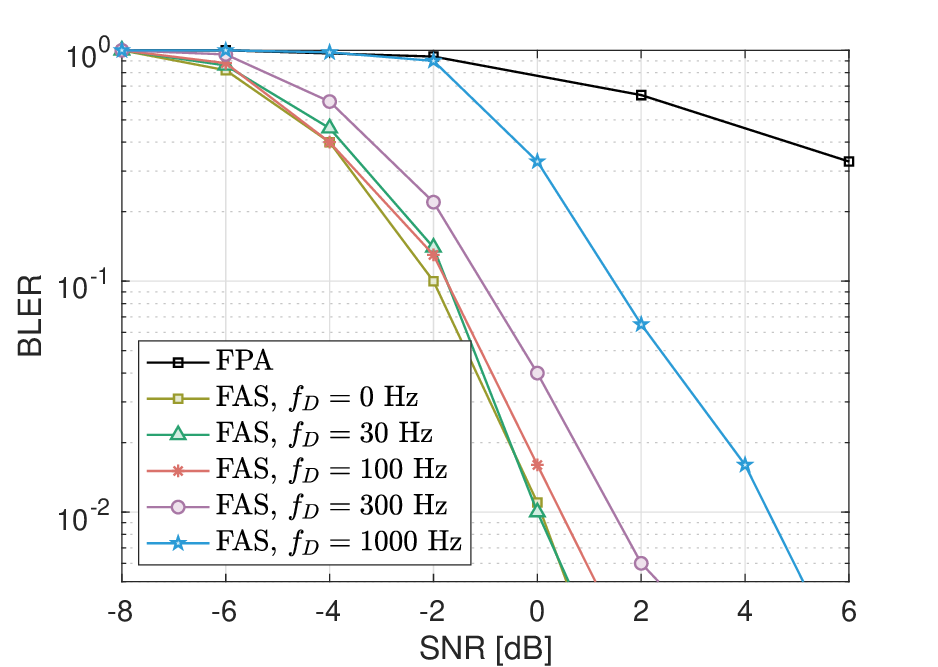}\label{SubFig:BLERvsSNR_DifffD_W1x1}}
\subfigure[$W = 5\lambda \times 5\lambda$]{\includegraphics[width=\linewidth]{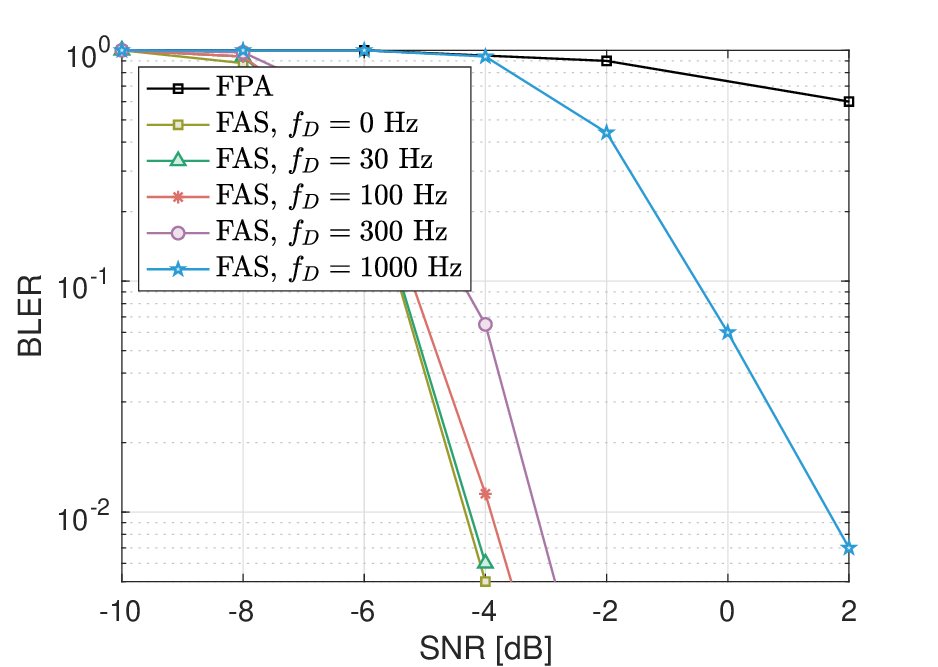}\label{SubFig:BLERvsSNR_DifffD_W5x5}}
\caption{BLER performance comparison for different delay spread. Results are presented for MCS7, with $K = 20\times 20$ antenna ports over the FAS with physical size of (a) $W = 1\lambda \times 1\lambda$ and (b) $W = 5\lambda \times 5\lambda$.}\label{Fig:BLERDifffD}
\end{figure}

In Fig.~\ref{Fig:BLERDifffD}, we provide the BLER results for various Doppler shift ($f_D$) values. Considering carrier frequency of $5$ GHz, a Doppler frequency of $f_D = 1000$ Hz corresponds to the relative velocity of $216$ km/h of the UE, which represents a very high-speed scenario. The maximum Doppler frequencies of $f_D = 300$ and $100$ Hz correspond to typical vehicular scenarios with relative velocities of $64.8$ and $21.6$ km/h, while $f_D = 30$ Hz indicates a pedestrian scenario with a velocity of $6.48$ km/h, and $f_D = 0$ Hz reflects a stationary scenario. The BLER performance demonstrates some decline in the mobile scenarios, as expected. The performance degradation observed in the pedestrian scenario is negligible, while in vehicular scenarios, it remains below $2$ dB. However, in the high-speed vehicular scenario, the performance degradation is more substantial, approximately $4.5$ dB for $W = 1\lambda \times 1\lambda$ and about $6$ dB for $W = 5\lambda \times 5\lambda$. Nevertheless, the performance remains much better than that of the FPA system. This notable performance degradation can be attributed to a low coherence time ($T_c = 1/f_D = 1~{\rm ms}$), which is equivalent to the duration of a subframe. As such, time-selective fading becomes severe, causing considerable performance degradation.

\subsection{AMC}\label{subsec:amc}
\begin{figure}
\centering
\includegraphics[width = \linewidth]{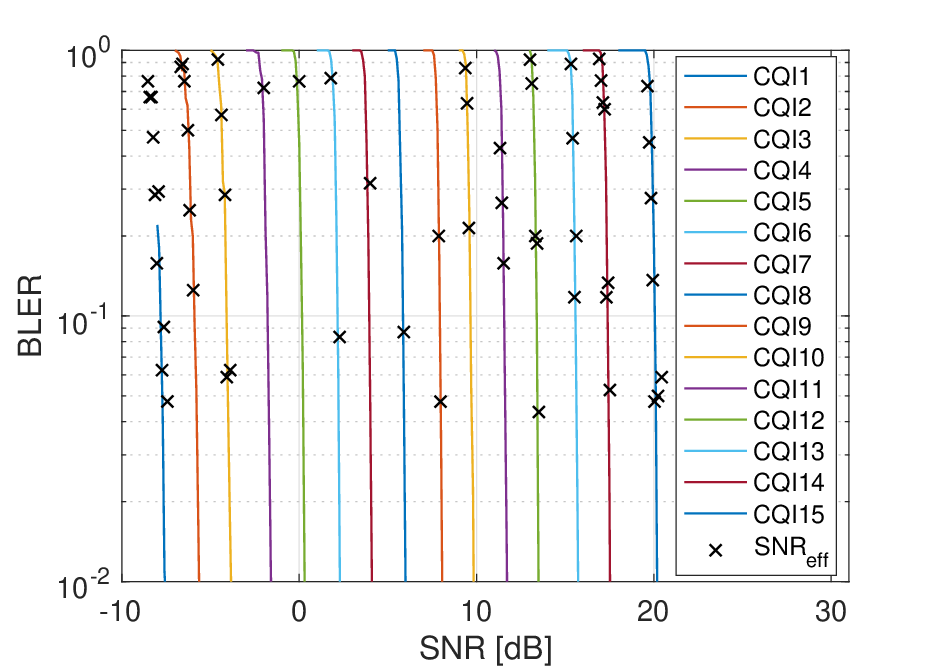}
\caption{Effective SNR mapping for AMC in the FAS-OFDM system with $K=20\times 20$ FAS antenna ports over the physical size of $W = 1\lambda \times 1\lambda$.}\label{Fig:EffSNRMap}
\end{figure}

As introduced in Section \ref{subsec:AMCMech}, we first conduct simulations of $15$ cases in the single-input single-output (SISO) system over AWGN channels, with results presented in Fig.~\ref{Fig:EffSNRMap}. By considering a criterion of $\text{BLER}=0.1$, the SNR threshold for the LUT (look-up table) can be deduced through linear fitting of the simulated results, as detailed in Table \ref{Tab:CQI} under the designation of $\Gamma_\text{LUT}$. Subsequently, simulations of FAS-OFDM are performed to ascertain the optimal adjustment factor $\alpha$. In this subsection, we consider the system featuring FAS with $K=20\times 20$ ports uniformly distributed over a physical size of $W = 1\lambda \times 1\lambda$. The adjustment factors related to the $15$ CQI indices are listed in the final column of Table \ref{Tab:CQI}. 

\begin{table}
\begin{center}
\caption{SNR Threshold and Optimal Adjustment Factor}\label{Tab:CQI}
\begin{tabular}{c|c|c|c|c}
\hline
\makecell[c]{\textbf{CQI}\\\textbf{index}}  & \textbf{Modulation}  & \makecell[c]{\textbf{Code Rate}\\$\times 1024$}   & \makecell[c]{\textbf{SNR threshold}\\$\Gamma_\text{LUT}$ (dB)}  & $\alpha$ \\ 
\hline\hline
    $0$   & \multicolumn{4}{c}{Out of range} \\ \hline
    $1$   & QPSK  & $78$    & $-7.84$ & $0.97$  \\ \hline
    $2$   & QPSK  & $120$   & $-5.94$ & $0.76$  \\ \hline
    $3$   & QPSK  & $193$   & $-4.11$ & $0.98$  \\ \hline
    $4$   & QPSK  & $308$  & $-1.80$ & $1.00$  \\ \hline
    $5$   & QPSK  & $449$   & $0.13$  & $1.00$  \\ \hline
    $6$   & QPSK  & $602$   & $2.16$  & $0.95$  \\ \hline\hline
    $7$   & 16QAM & $378$   & $3.93$  & $1.00$  \\ \hline
    $8$   & 16QAM & $490$   & $5.83$  & $1.00$  \\ \hline
    $9$   & 16QAM & $616$   & $7.93$  & $0.97$  \\ \hline\hline
    $10$  & 64QAM & $466$   & $9.67$  & $0.97$ \\ \hline
    $11$  & 64QAM & $567$   & $11.56$ & $0.96$ \\ \hline
    $12$  & 64QAM & $666$   & $13.38$ & $0.98$ \\ \hline
    $13$  & 64QAM & $772$   & $15.57$ & $0.96$ \\ \hline
    $14$  & 64QAM & $873$   & $17.39$ & $0.96$ \\ \hline
    $15$  & 64QAM & $948$   & $20.03$ & $0.92$ \\ \hline
\end{tabular}
\end{center}
\end{table}

Fig.~\ref{Fig:EffSNRMap} presents the effective SNR mapping for FAS-OFDM operating under the TDL-C channel, characterized by $\text{DS}=30$ ns and $f_D = 30$ Hz. The solid lines represent the SNR-BLER curves for the SISO system over the AWGN channel, whereas the black `X' markers depict the BLER against the effective SNR as mapped by \eqref{Eq:effSNR}, using the adjustment factors $\alpha$ in Table \ref{Tab:CQI}. The `X' points in Fig.~\ref{Fig:EffSNRMap} are relatively concentrated on the AWGN curves, which indicates that the AMC, supported by the effective SNR mapping in \eqref{Eq:effSNR}, is highly accurate and effectively reflects the channel quality of FAS-OFDM.  

\subsection{Throughput}\label{subsec:SE}
\begin{figure}[]
\centering
\subfigure[$\text{BW} = 5$ MHz]{\includegraphics[width = \linewidth]{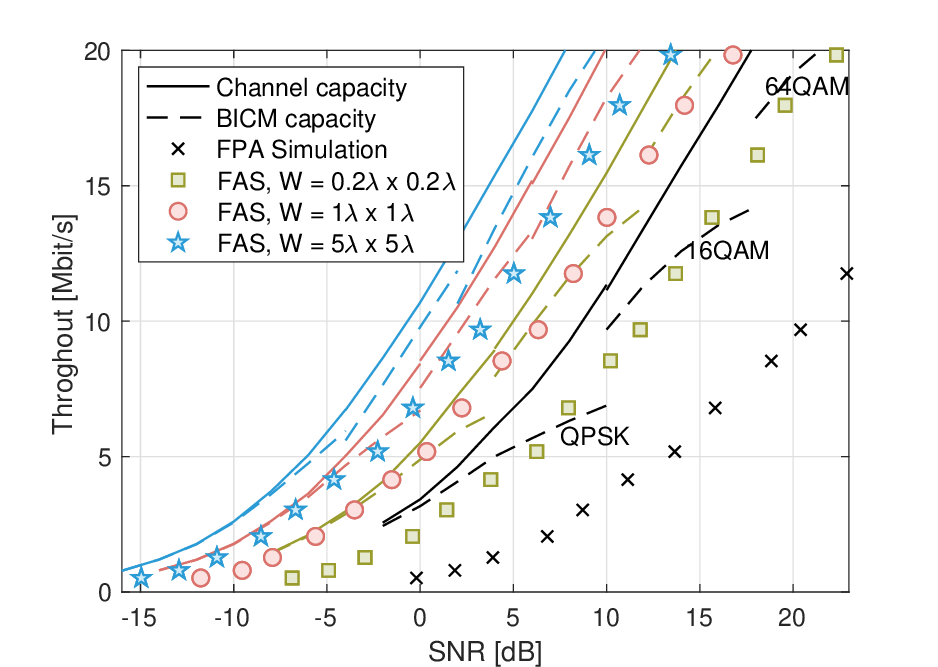}\label{SubFig:throughput_BW5MHz}}
\subfigure[$\text{BW} = 20$ MHz]{\includegraphics[width = \linewidth]{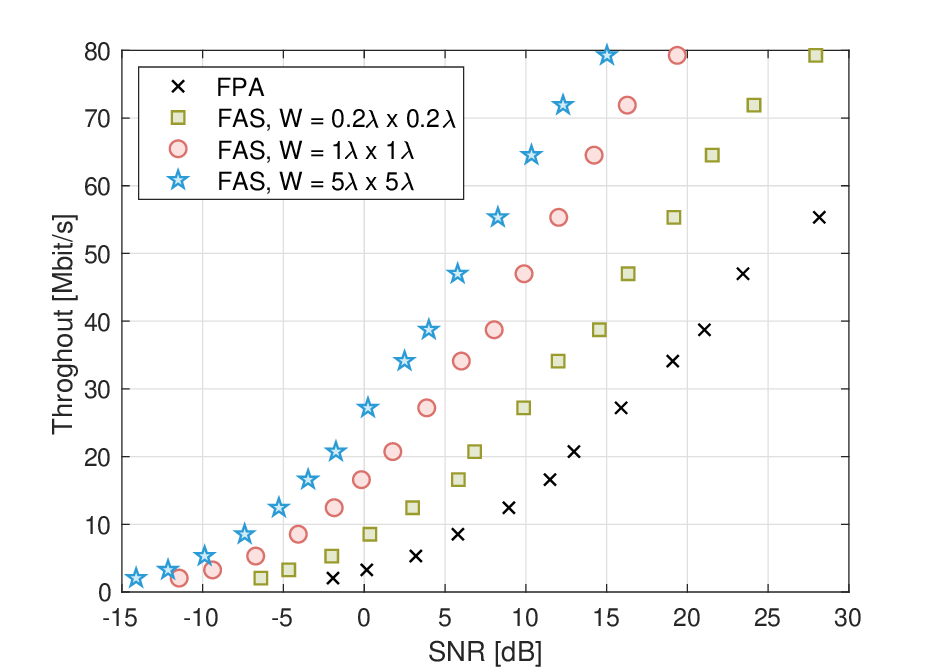}\label{SubFig:throughput_BW20MHz}}
\caption{Throughput against SNR in dB of the FAS-OFDM system. Results are presented for (a) $5$ MHz bandwidth and (b) $20$ MHz bandwidth.}\label{Fig:throughput}
\end{figure}

Fig.~\ref{Fig:throughput} presents the throughput of FAS-OFDM using the $15$ CQI indices. The results are based on the TDL channel, with $\text{DS} = 30$ ns and $f_D = 30$ Hz. The results depicted in this figure are based on the criterion of $\text{BLER} = 0.1$. The SNR is the transmission SNR when $\text{BLER} = 0.1$, and the throughput is calculated as $\gamma \times (1-\text{BLER})$, where $\gamma$ is the overall system throughput given in \eqref{Eq:throughput}. Simulations for the FPA system are also conducted for comparison. For the FAS-OFDM system, the FAS configuration involves $K = 20 \times 20$ ports uniformly distributed over varying sizes: $W = 0.2\lambda \times 0.2\lambda$, $1\lambda \times 1\lambda$, and $5\lambda \times 5\lambda$. Additionally, the channel capacity and the BICM capacity are also plotted for comparison within the $5$ MHz bandwidth system in Fig. \ref{Fig:throughput}\ref{sub@SubFig:throughput_BW5MHz}. These two rates are calculated in accordance with Section \ref{subsec:AR}, with adjustments made to account for the RS, CP, and GB overheads, as specified in \eqref{Eq:rela_E_EBICM}. The BICM capacity calculations utilize QPSK, 16QAM, and 64QAM in low, medium, and high SNR region, respectively.

As expected, the simulation results are upper-bounded by the semi-analytical capacity results. Note that the gaps between the simulations and the semi-analytical results are minimal in the FAS-OFDM system, especially when the size of FAS is large. This observation may indicate that FAS effectively approaches the capacity. Moreover, the throughput in the FAS-OFDM system shows a considerable improvement over that of the FPA system, with more throughput gain as the size of FAS increases. A commonly used SNR value of interest for reception is $10$ dB. Under this particular condition, the FPA system provides throughput of $3.62$ Mbit/s and $14.15$ Mbit/s for the $5$ MHz and $20$ MHz bandwidth, respectively, corresponding to a spectral efficiency of $0.7$ bit/s/Hz. Conversely, the FAS-OFDM system achieves throughput of $8.39$ Mbit/s and $27.63$ Mbit/s, even with a relatively small FAS size $W = 0.2\lambda \times 0.2\lambda$, effectively doubling the throughput of the FPA system. With a configuration of $W = 1\lambda \times 1\lambda$, the throughput increases to $13.81$ Mbit/s for $5$ MHz bandwidth and $47.43$ Mbit/s for $20$ MHz bandwidth. This performance further escalates, reaching $17.18$ Mbit/s and $62.97$ Mbit/s with the $W = 5\lambda \times 5\lambda$ FAS configuration. Overall, the FAS-OFDM system significantly enhances throughput, achieving improvements ranging from $95\%$ to $375\%$ compared to the FPA system.

\section{Conclusion}\label{sec:conclusion}
In this paper, we integrated FAS within the OFDM framework to address the challenges of utilizing FAS for wideband communications. We proposed a comprehensive framework for FAS-OFDM that includes channel-aware port selection metrics and the AMC mechanism. The proposed system utilizes the PDSCH procedures in the 5G NR specifications. The link-level simulation results showed great improvements in the BLER and throughput, even in a small space, highlighting the great potential of FAS for future wireless communications.   

\bibliographystyle{IEEEtran}
%  \bibliography{Reference}

\begin{thebibliography}{10}
\providecommand{\url}[1]{#1}
\csname url@samestyle\endcsname
\providecommand{\newblock}{\relax}
\providecommand{\bibinfo}[2]{#2}
\providecommand{\BIBentrySTDinterwordspacing}{\spaceskip=0pt\relax}
\providecommand{\BIBentryALTinterwordstretchfactor}{4}
\providecommand{\BIBentryALTinterwordspacing}{\spaceskip=\fontdimen2\font plus
\BIBentryALTinterwordstretchfactor\fontdimen3\font minus
  \fontdimen4\font\relax}
\providecommand{\BIBforeignlanguage}[2]{{%
\expandafter\ifx\csname l@#1\endcsname\relax
\typeout{** WARNING: IEEEtran.bst: No hyphenation pattern has been}%
\typeout{** loaded for the language `#1'. Using the pattern for}%
\typeout{** the default language instead.}%
\else
\language=\csname l@#1\endcsname
\fi
#2}}
\providecommand{\BIBdecl}{\relax}
\BIBdecl
\bibitem{Ericsson6G}
Ericsson, ``Co-creating a cyber-physical world,'' Available [Online]: \url{https://www.ericsson.com/en/reports-and-papers/white-papers/co-creating-a-cyber-physical-world}, Last Accessed on 2024-08-12.
\bibitem{Andrews20246G}
J. G. Andrews, T. E. Humphreys and T. Ji, ``6G Takes Shape,'' {\em IEEE BITS the Information Theory Magazine}, early access, \url{doi: 10.1109/MBITS.2024.3504521}, Nov. 2024.
\bibitem{8766143}
Z.~Zhang {\em et al.}, ``{6G} wireless networks: Vision, requirements, architecture, and key technologies,'' {\em IEEE Veh. Technol. Mag.}, vol.~14, no.~3, pp.~28--41, Sept. 2019.
\bibitem{8869705}
W.~Saad, M.~Bennis, and M.~Chen, ``A vision of {6G} wireless systems: Applications, trends, technologies, and open research problems,'' {\em IEEE Netw.}, vol.~34, no.~3, pp.~134--142, May/Jun. 2020.
\bibitem{Tariq-2020}
F.~Tariq {\em et al.}, ``A speculative study on {6G},'' {\em IEEE Wireless Commun.}, vol.~27, no.~4, pp.~118--125, Aug. 2020.
\bibitem{10054381}
C.-X. Wang {\em et al.}, ``On the road to {6G}: Visions, requirements, key technologies, and testbeds,'' {\em IEEE Commun.  Surv. \& Tut.}, vol.~25, no.~2, pp.~905--974, Secondquarter 2023.

\bibitem{wong2020FAS}
K. K. Wong, K. F. Tong, Y. Zhang, and Z. Zheng, ``Fluid antenna system for {6G}: When {Bruce Lee} inspires wireless communications,'' {\em Elect. Lett.}, vol.~56, no.~24, pp.~1288--1290, Nov. 2020.
\bibitem{wong2022bruce}
K.-K. Wong, K.-F. Tong, Y.~Shen, Y.~Chen, and Y.~Zhang, ``{Bruce Lee}-inspired fluid antenna system: Six research topics and the potentials for {6G},'' {\em Front. Commun. Netw. Sec. Wireless Commun.}, vol. 3, Mar. 2022.
\bibitem{Wang-2024ai}
C. Wang {\em et al.}, ``AI-empowered fluid antenna systems: Opportunities, challenges, and future directions,'' {\em IEEE Wireless Commun.}, vol. 31, no. 5, pp. 34--41, Oct. 2024.
\bibitem{wu2024fluid}
T.~Wu {\em et al.}, ``Fluid antenna systems enabling {6G}: Principles, applications, and research directions,'' {\em arXiv preprint}, \url{arXiv:2412.03839}, 2024.
\bibitem{New2024aTutorial}
W. K. New {\em et al.}, ``A tutorial on fluid antenna system for 6G networks: Encompassing communication theory, optimization methods and hardware designs,'' \emph{IEEE Commun. Surv. \& Tutor.}, early access, \url{doi: 10.1109/COMST.2024.3498855}, 2024.
\bibitem{Lu-2025}
W.-J. Lu {\em et al.}, ``Fluid antennas: Reshaping intrinsic properties for flexible radiation characteristics in intelligent wireless networks,'' accepted in {\em IEEE Commun. Mag.}, 2025.
\bibitem{zhu2024historical} 
L.~Zhu and K. K. Wong, ``Historical review of fluid antenna and movable antenna,'' {\em arXiv preprint}, \url{arXiv:2401.02362v2}, 2024.
\bibitem{10480333}
J.~Zheng {\em et al.},  ``Flexible-position {MIMO} for wireless communications: Fundamentals, challenges, and future directions,'' {\em IEEE Wireless Commun.}, vol.~31, no.~5, pp.~18--26, Oct. 2024.
\bibitem{Yang-2025pa}
Z. Yang {\em et al.}, ``Pinching antennas: Principles, applications and challenges,'' {\em arXiv preprint}, \url{arXiv:2501.10753}, 2025.

\bibitem{9131873}
K.~K. Wong, A.~Shojaeifard, K.-F. Tong, and Y.~Zhang, ``Performance limits of fluid antenna systems,'' {\em IEEE Commun. Lett.}, vol.~24, no.~11, pp.~2469--2472, Nov. 2020.
\bibitem{wong2021FAS}
K. K. Wong, A. Shojaeifard, K. F. Tong, and Y. Zhang, ``Fluid antenna systems,'' {\em IEEE Trans. Wireless Commun.}, vol. 20, no. 3, pp. 1950--1962, Mar. 2021.

% FAS implementation
\bibitem{huang2021liquid}
Y. Huang, L. Xing, C. Song, S. Wang and F. Elhouni, ``Liquid antennas: Past, present and future,'' {\em IEEE Open J. Antennas \& Propag.}, vol.~2, pp. 473--487, Mar. 2021.
\bibitem{shen2024design}
Y. Shen {\em et al.}, ``Design and implementation of mmWave surface wave enabled fluid antennas and experimental results for fluid antenna multiple access,'' {\em arXiv preprint}, \url{arXiv:2405.09663}, May 2024.
\bibitem{Shamim-2025}
R. Wang {\em et al.}, ``Electromagnetically reconfigurable fluid antenna system for wireless communications: Design, modeling, algorithm, fabrication, and experiment,'' {\em arXiv preprint}, \url{arXiv:2502.19643v2}, 2025.
\bibitem{basbug2017design}
S. Basbug, ``Design and synthesis of antenna array with movable elements along semicircular paths,'' {\em IEEE Antennas Wireless Propag. Lett.}, vol.~16, pp. 3059--3062, Oct. 2017.
\bibitem{johnson2015sidelobe}
M. C. Johnson, S. L. Brunton, N. B. Kundtz, and J. N. Kutz, ``Sidelobe canceling for reconfigurable holographic metamaterial antenna,'' {\em IEEE Trans. Antennas \& Propag.}, vol.~63, no.~4, pp.~1881--1886, Apr. 2015.
\bibitem{hoang2021computational}
T. V. Hoang, V. Fusco, T. Fromenteze and O. Yurduseven, ``Computational polarimetric imaging using two-dimensional dynamic metasurface apertures,'' {\em IEEE Open J. Antennas \& Propag.}, vol.~2, pp. 488--497, Mar. 2021.
\bibitem{Liu-2025arxiv}
B. Liu, K. F. Tong, K. K. Wong, C.-B. Chae, and H. Wong, ``Be water, my antennas: Riding on radio wave fluctuation in nature for spatial multiplexing using programmable meta-fluid antenna,'' {\em arXiv preprint}, \url{arXiv:2502.04693}, 2025.
\bibitem{zhang2024pixel}
J. Zhang {\em et al.}, ``A novel pixel-based reconfigurable antenna applied in fluid antenna systems with high switching speed,'' {\em IEEE Open J. Antennas \& Propag.}, early access, \url{doi: 10.1109/OJAP.2024.3489215}, 2024.

% FAS different channel
\bibitem{Khammassi2023}
M. Khammassi, A. Kammoun and M.-S. Alouini, ``A new analytical approximation of the fluid antenna system channel,'' {\em IEEE Trans. Wireless Commun.}, vol. 22, no. 12, pp. 8843--8858, Dec. 2023.
\bibitem{New2023fluid}
W. K. New, K. K. Wong, H. Xu, K. F. Tong and C.-B. Chae, ``Fluid antenna system: New insights on outage probability and diversity gain,''  {\em IEEE Trans. Wireless Commun.}, vol. 23, no. 1, pp. 128--140, Jan. 2024.
\bibitem{Espinosa-2024}
P. Ram\'{i}rez-Espinosa, D. Morales-Jimenez, and K. K. Wong, ``A new spatial block-correlation model for fluid antenna systems,'' {\em IEEE Trans. Wireless Commun.}, vol. 23, no. 11, pp. 15829--15843, Nov. 2024.
\bibitem{Vega2023asimple}
J. D. Vega-S$\acute{\rm a}$nchez, L. Urquiza-Aguiar, M. C. P. Paredes, and D. P. M. Osorio, ``A simple method for the performance analysis of fluid antenna systems under correlated Nakagami-$m$ fading,'' {\em IEEE Wireless Commun. Lett.}, vol. 13, no. 2, pp. 377--381, Feb. 2024.
\bibitem{Vega2023novel}
J. D. Vega-S$\acute{\rm a}$nchez, A. E. L$\acute{\rm o}$pez-Ram$\acute{\rm i}$rez, L. Urquiza-Aguiar, and D. P. M. Osorio, ``Novel expressions for the outage probability and diversity gains in fluid antenna system,'' {\em IEEE Wireless Commun. Lett.}, vol. 13, no. 2, pp. 372--376, Feb. 2024.
\bibitem{Alvim2023on}
P. D. Alvim {\em et al.}, ``On the performance of fluid antennas systems under $\alpha$-$\mu$ fading channels,'' {\em IEEE Wireless Commun. Lett.}, vol. 13, no. 1, pp. 108--112, Jan. 2024.
\bibitem{new2023information}
W. K. New, K. K. Wong, H. Xu, K. F. Tong, and C.-B. Chae, ``An information-theoretic characterization of {MIMO-FAS}: Optimization, diversity-multiplexing tradeoff and \textit{q}-outage capacity,'' \emph{IEEE Trans. Wireless Commun.}, vol. 23, no. 6, pp. 5541--5556, Jun. 2024.

% FAS channel estimation
\bibitem{Skouroumounis2023fluid}
C. Skouroumounis and I. Krikidis, ``Fluid antenna with linear MMSE channel estimation for large-scale cellular networks,'' {\em IEEE Trans. Commun.}, vol. 71, no. 2, pp. 1112--1125, Feb. 2023.
\bibitem{xu2024channel}
H. Xu {\em et al.}, ``Channel estimation for FAS-assisted multiuser mmWave systems,'' {\em IEEE Commun. Lett.}, vol. 28, no. 3, pp. 632--636, Mar. 2024.
\bibitem{zhang2023successive}
Z. Zhang, J. Zhu, L. Dai, and R. W. Heath Jr, ``Successive Bayesian reconstructor for channel estimation in fluid antenna systems,'' {\em IEEE Trans. Wireless Commun.}, \url{doi: 10.1109/TWC.2024.3515135}, 2024.
\bibitem{Xu-2025ce}
B. Xu, Y. Chen, Q. Cui, X. Tao, and K. K. Wong, ``Sparse Bayesian learning-based channel estimation for fluid antenna systems,'' {\em IEEE Wireless Commun. Lett.}, vol. 14, no. 2, pp. 325--329, Feb. 2025.
\bibitem{10751774}
W.~K. New {\em et al.}, ``Channel estimation and reconstruction in fluid antenna system: Oversampling is essential,'' {\em IEEE Trans. Wireless Commun.}, vol. 24, no. 1, pp. 309--322, Jan. 2025.

% FAS application
\bibitem{Tang-2023}
B. Tang {\em et al.}, ``Fluid antenna enabling secret communications,'' {\em IEEE Commun. Lett.}. vol. 27, no. 6, pp. 1491--1495, Jun. 2023.
\bibitem{Xu-2024pls}
H. Xu {\em et al.}, ``Coding-enhanced cooperative jamming for secret communication in fluid antenna systems,'' {\em IEEE Commun. Lett.}, vol. 28, no. 9, pp. 1991--1995, Sept. 2024.
\bibitem{Ghadi-2024dec}
F. R. Ghadi {\em et al.}, ``Physical layer security over fluid antenna systems: Secrecy performance analysis,'' {\em IEEE Trans. Wireless Commun.}, vol. 23, no. 12, pp. 18201--18213, Dec. 2024.
\bibitem{10539238}
F.~Rostami~Ghadi {\em et al.}, ``On performance of {RIS}-aided fluid antenna systems,'' {\em IEEE Wireless Commun. Lett.}, vol.~13, no.~8, pp.~2175--2179, Aug. 2024.
\bibitem{Zhu-2025ris}
J. Zhu {\em et al.}, ``Fluid antenna empowered index modulation for RIS-aided mmWave transmissions,'' {\em IEEE Trans. Wireless Commun.}, vol. 24, no. 2, pp. 1635--1647, Feb. 2025.
\bibitem{Yao2025RIS}
J. Yao {\em et al.}, ``FAS-RIS communication: Model, analysis, and optimization,'' {\em IEEE Trans. Veh. Technol.}, \url{doi:10.1109/TVT.2025.3537294}, 2025.
\bibitem{Salem-2025ris}
A. Salem, K. K. Wong, G. Alexandropoulos, C.-B. Chae, and R. Murch, ``A first look at the performance enhancement potential of fluid reconfigurable intelligent surface,'' {\em arXiv preprint}, \url{arXiv:2502.17116v1}, 2025.
\bibitem{Wang-2024isac}
C. Wang {\em et al.}, ``Fluid antenna system liberating multiuser MIMO for ISAC via deep reinforcement learning,'' {\em IEEE Trans. Wireless Commun.}, vol. 23, no. 9, pp. 10879--10894, Sept. 2024.
\bibitem{zhou2024fasisac}
L. Zhou, J. Yao, M. Jin, T. Wu and K. K. Wong, ``Fluid antenna-assisted ISAC systems,'' \emph{IEEE Wireless Commun. Lett.}, vol. 13, no. 12, pp. 3533--3537, Dec. 2024.
\bibitem{Zou-2024}
J. Zou {\em et al.}, ``Shifting the ISAC trade-off with fluid antenna systems,'' {\em IEEE Wireless Commun. Lett.}, vol. 13, no. 12, pp. 3479--3483, Dec. 2024.

% FAMA
\bibitem{wong2022FAMA}
K. K. Wong and K. F. Tong, ``Fluid antenna multiple access,'' \emph{IEEE Trans. Wireless Commun.}, vol.~21, no.~7, pp. 4801--4815, Jul. 2022.
\bibitem{wong2023sFAMA}
K. K. Wong, D. Morales-Jimenez, K. F. Tong, and C. B. Chae, ``Slow fluid antenna multiple access,'' \emph{IEEE Trans. Commun.}, vol.~71, no.~5, pp. 2831--2846, May 2023.
\bibitem{Xu2024revisiting}
H.~Xu {\em et al.}, ``Revisiting outage probability analysis for two-user fluid antenna multiple access system,'' \emph{IEEE Trans. Wireless Commun.}, vol. 23, no. 8, pp. 9534--9548, Aug. 2024.
\bibitem{Waqar-2023}
N. Waqar, K. K. Wong, K. F. Tong, A. Sharples, and Y. Zhang, ``Deep learning enabled slow fluid antenna multiple access,'' {\em IEEE Commun. Lett.}, vol. 27, no. 3, pp. 861--865, Mar. 2023. 
\bibitem{hong2024coded}
H. Hong, K. K. Wong, K. F. Tong, H. Shin, and Y. Zhang, ``Coded fluid antenna multiple access over fast fading channels,'' \emph{IEEE Wireless Commun. Lett.}, \url{doi:10.1109/LWC.2025.3540668}, 2025.
\bibitem{hong2025Downlink}
H. Hong, K. K. Wong, H. Xu, {\em et al.}, ``Downlink OFDM-FAMA in 5G-NR Systems,'' {\em arXiv preprint}, \url{arxiv:2501.06974}, Jan. 2025.
\bibitem{Wong2024cuma}
K. K. Wong, C. B. Chae, and K. F. Tong, ``Compact ultra massive antenna array: A simple open-loop massive connectivity scheme,'' {\em IEEE Trans. Wireless Commun.}, vol. 23, no. 6, pp. 6279--6294, Jun. 2024.
\bibitem{Wong-2024cuma-ris}
K. K. Wong, ``Transmitter CSI-free RIS-randomized CUMA for extreme massive connectivity,'' {\em IEEE Open J. Commun. Soc.}, vol. 5, pp. 6890--6902, 2024.

% 3GPP TR/TS
\bibitem{38214}
``{NR}; {P}hysical layer procedures for data,'' Available [Online]: \url{https://www.3gpp.org/ftp/Specs/archive/38_series/38.214/38214-i40.zip}, Last Accessed on 2024-09-23.
\bibitem{38212}
``{NR}; {M}ultiplexing and channel coding,'' Available [Online]: \url{https://www.3gpp.org/ftp/Specs/archive/38_series/38.212/38212-i40.zip}, Last Accessed on 2024-09-23.
\bibitem{38211}
``{NR}; {P}hysical channels and modulation,'' Available [Online]: \url{https://www.3gpp.org/ftp/Specs/archive/38_series/38.211/38211-i40.zip}, Last Accessed on 2024-09-23.
\bibitem{38901}
``Study on channel model for frequencies from 0.5 to 100 GHz,'' Available [Online]: \url{https://www.3gpp.org/ftp/Specs/archive/38~series/38.901/8901-i00.zip}, Last Accessed on 2024-04-03.
\bibitem{BICM}
G. Caire, G. Taricco and E. Biglieri, ``Bit-interleaved coded modulation,'' {\em IEEE Trans. Inf. Theory}, vol.~44, no.~3, pp. 927--946, May 1998.
\end{thebibliography}

\end{document}